\pdfoutput=1
\documentclass[11pt,twocolumn]{article}

\usepackage{times}
\usepackage[
    left=1.75cm,
    right=1.75cm,
    top=3.00cm,
    bottom=3.00cm
]{geometry}

\usepackage[normalem]{ulem}
\usepackage[abbreviations]{foreign}  
\usepackage{fdsymbol} 
\usepackage{booktabs} 
\usepackage[T1]{fontenc} 
\usepackage{pifont}
\usepackage{listings}
\usepackage{xspace}
\usepackage{setspace}
\usepackage{enumitem}
\usepackage{framed}
\usepackage[usenames, dvipsnames]{xcolor}
\usepackage{framed}
\definecolor{shadecolor}{RGB}{240,240,240}
\usepackage{graphicx} 
\usepackage{caption}
\usepackage{url}

\usepackage{breakurl}
\usepackage[]{hyperref}
\hypersetup{
  colorlinks,
  linkcolor={green!50!black},
  citecolor={red!70!black},
  urlcolor={blue!70!black}
}

\usepackage[acronym]{glossaries}
\newacronym{sgx}{SGX}{Intel software guard extensions}
\newacronym{txt}{TXT}{Intel trusted execution technology}
\newacronym{sem}{AMD SEM}{AMD secure execution mode}
\newacronym{tee}{TEE}{trusted execution environment}
\newacronym{kms}{KMS}{key management system}
\newacronym{epc}{EPC}{enclave page cache}
\newacronym{tls}{TLS}{transport layer security}
\newacronym{os}{OS}{operating system}
\newacronym{drtm}{DRTM}{dynamic root of trust for measurements}
\newacronym{srtm}{SRTM}{static root of trust for measurements}
\newacronym{crtm}{CRTM}{core root of trust for measurements}
\newacronym{rom}{ROM}{read-only memory}
\newacronym{nvram}{NVRAM}{non-volatile RAM}
\newacronym{dram}{DRAM}{dynamic random-access memory}
\newacronym{ek}{EK}{endorsement key}
\newacronym{aik}{AIK}{attestation key}
\newacronym{ca}{CA}{certificate authority}
\newacronym{tpm}{TPM}{trusted platform module}
\newacronym{dtpm}{dTPM}{discrete TPM chip}
\newacronym{ftpm}{fTPM}{firmware TPM}
\newacronym{itpm}{iTPM}{integrated TPM}
\newacronym{pch}{PCH}{platform controller hub}
\newacronym{tcg}{TCG}{Trusted Computer Group}
\newacronym{pcr}{PCR}{platform configuration register}
\newacronym{ptt}{Intel PTT}{Intel platform trusted technology}
\newacronym{uefi}{UEFI}{unified extensible firmware interface}
\newacronym{bios}{BIOS}{basic input/output system}
\newacronym{pxe}{PXE}{preboot execution environment}
\newacronym{svm}{SVM}{Secure Virtual Machine}
\newacronym{ima}{IMA}{integrity measurement architecture}
\newacronym{vpn}{VPN}{virtual private network}
\newacronym{daa}{DAA}{direct anonymous attestation}
\newacronym{loc}{LOC}{lines of code}
\newacronym{mee}{MEE}{memory encryption engine}
\newacronym{ias}{IAS}{Intel Attestation Service}
\newacronym{mrenclave}{MRENCLAVE}{enclave hash measurement}
\newacronym{acs}{IBM ACS}{IBM TPM attestation client-server}
\newacronym{cit}{Intel CIT}{Intel open cloud integrity technology}
\newacronym{secl}{Intel SECL-DC}{Intel security libraries for data center}
\newacronym{dc}{DC}{data center}
\newglossaryentry{initramfs}{name={initramfs},description={}}
\newacronym{vm}{VM}{virtual machine}
\newacronym{iaas}{IaaS}{Infrastructure-as-a-Service}
\newacronym{caas}{CaaS}{Container-as-a-Service}
\newacronym{maas}{MaaS}{Metal-as-a-Service}
\newacronym{lpc}{LPC}{low pin count}
\newacronym{me}{CSME}{Intel converged security and manageability engine}
\newacronym{toctou}{TOCTOU}{time of check to time of use}
\newacronym{itl}{ITL}{Invisible Things Lab}
\newacronym{smm}{SMM}{system management mode}
\newacronym{dma}{DMA}{direct memory access}
\newacronym{tcb}{TCB}{trusted computing base}
\newacronym{bmc}{BMC}{baseboard management controller}
\newacronym{ipmi}{IPMI}{intelligent platform management interface}
\newacronym{nic}{NIC}{network interface card}
\newacronym{rest}{REST}{representational state transfer}
\newacronym{gps}{GPS}{global positioning system}
\newacronym{api}{API}{application programming interface}
\newacronym{fpga}{FPGA}{field-programmable gate array}
\newacronym{cpu}{CPU}{central processing unit}
\newacronym{arp}{ARP}{address resolution protocol}
\newacronym{tcp}{TCP}{transmission control protocol}
\newacronym{dns}{DNS}{domain name system}
\newacronym{gdpr}{GDPR}{general data protection regulation}
\newacronym{tct}{TCT}{trusted computing technologies}
\newacronym{tbe}{agent}{}
\newacronym{erezept}{erezept}{TOFIX?}
\newacronym{hmac}{HMAC}{hash-based message authentication code}
\newacronym{rsa}{RSA}{Rivest–Shamir–Adleman}
\newacronym{ecdsa}{ECDSA}{elliptic curve digital signature algorithm}
\newacronym{ml}{ML}{machine learning}
\newacronym{siem}{SIEM}{security information and event management}

\glsdisablehyper
\makeglossaries

\newcommand\YAMLcolonstyle{\color{red!70!black}}
\newcommand\YAMLkeystyle{\color{black}}
\newcommand\YAMLvaluestyle{\color{blue!70!black}}
\newcommand\YAMLcommentstyle{\color{blue!50!black}}

\makeatletter

\newcommand\language@yaml{yaml}
\expandafter\expandafter\expandafter\lstdefinelanguage
\expandafter{\language@yaml}
{
	keywords={true,false,null,y,n},
	keywordstyle=\color{darkgray}\bfseries,
    basicstyle=\linespread{1}\footnotesize\ttfamily\color{black},
	sensitive=false,
	comment=[l]{\#},
	morecomment=[s]{/*}{*/},
	commentstyle=\YAMLcommentstyle,
	stringstyle=\YAMLvaluestyle,
	moredelim=[l][\color{orange}]{\&},
	moredelim=[l][\color{magenta}]{*},
	moredelim=**[il][\YAMLcolonstyle{:}\YAMLvaluestyle]{},
	morestring=[b]',
	morestring=[b]",
	literate = {---}{{\ProcessThreeDashes}}3
	{>}{{\textcolor{red!70!black}\textgreater}}1
	{|}{{\textcolor{red!70!black}\textbar}}1
	{-}{{\mdseries-}}1,
}

\lst@AddToHook{EveryLine}{\ifx\lst@language\language@yaml\YAMLkeystyle\fi}
\makeatother

\newcommand\SAPICcolonstyle{\color{red!70!black}}
\newcommand\SAPICkeystyle{\color{black}}
\newcommand\SAPICvaluestyle{\color{blue!70!black}}
\newcommand\SAPICcommentstyle{\color{blue!50!black}\scriptsize}

\makeatletter

\newcommand\language@sapic{sapic}

\expandafter\expandafter\expandafter\lstdefinelanguage
\expandafter{\language@sapic}
{
	keywords={true,false,null,y,n,in,out,new,if,then,else,event,pk, sign,senc,hash, verify},
	keywordstyle=\color{black}\bfseries,
	basicstyle=\linespread{1}\scriptsize\ttfamily\color{black},
	sensitive=false,
	comment=[l]{//},	
	morecomment=[s]{/*}{*/},
	commentstyle=\SAPICcommentstyle,
	stringstyle=\SAPICvaluestyle,
	moredelim=[l][\color{orange}]{\&},
	moredelim=[l][\color{magenta}]{*},
	moredelim=**[il][\SAPICcolonstyle{:}\SAPICvaluestyle]{},
	morestring=[b]',
	morestring=[b]",
	literate = {---}{{\ProcessThreeDashes}}3
	{<}{{\textcolor{red!70!black}\textless}}1
	{>}{{\textcolor{red!70!black}\textgreater}}1
	{|}{{\textcolor{red!70!black}\textbar}}1
	{!}{{\textcolor{red!70!black}{\string!}}}1
	{-}{{\mdseries-}}1,
}

\lst@AddToHook{EveryLine}{\ifx\lst@language\language@sapic\SAPICkeystyle\fi}
\makeatother

\lstdefinelanguage{Ini}
{
	basicstyle=\scriptsize\ttfamily,
	morecomment=[s][\color{blue}]{[}{]},
	morecomment=[l]{;},
	commentstyle=\color{gray},
	morekeywords={},
	otherkeywords={=},
	keywordstyle={\color{green}}
}

\newcommand{\sys}{\textsc{Synergía}\xspace}
\newcommand{\tpmquote}{quote\xspace}
\newcommand{\imalog}{IMA log\xspace}
\newcommand{\imapcr}{IMA PCR\xspace}

\newcommand{\sysinit}{agent initialization\xspace}
\newcommand{\sysruntime}{agent runtime\xspace}

\newcommand{\ubuntuver}{Ubuntu\,16.04 LTS\xspace}
\newcommand{\kernelver}{Linux kernel\,v4.4.0-135-generic\xspace}
\newcommand{\machinetype}{Dell PowerEdge\,R330\xspace}
\newcommand{\machinecpu}{Intel Xeon E3-1270\,v5 CPU\xspace}
\newcommand{\tpmchipversion}{Infineon\,9665\,TPM 2.0}
\newcommand{\microcodever}{0xc6\xspace}

\newcommand{\cuckooattack}{cuckoo attack\xspace}

\newcommand{\myparagraph}[1]{\vspace{1mm} \smallskip \noindent{\bf {#1}.}}
\newcommand{\myparagraphnotdot}[1]{\vspace{1mm} \smallskip \noindent{\bf {#1}}}

\let\origthelstnumber\thelstnumber
\makeatletter
\renewcommand\thelstnumber{%
    \ifnum\value{lstnumber}>0
        \origthelstnumber
    \else
        \ifnum\value{lstnumber}=-1
            \ldots
        \fi
    \fi
}
\newcommand*\startnumber[1]{%
    \setcounter{lstnumber}{\numexpr#1-1\relax}%
}
\newcommand*\stopnumber{%
    \startnumber{-2}%
}
\makeatother

\AtBeginDocument{%
  \providecommand\BibTeX{{%
    \normalfont B\kern-0.5em{\scshape i\kern-0.25em b}\kern-0.8em\TeX}}}

\begin{document}
\title{\sys: Hardening High-Assurance Security Systems with Confidential and Trusted Computing}

\author{
        Wojciech Ozga\\
        \textit{IBM Research -- Zurich} \\
        \textit{TU Dresden}
    \and
    Rasha Faqeh\\
        \textit{TU Dresden}        
        \and
        Do Le Quoc\\
        \textit{Huawei Research}
    \and
        Franz Gregor\\
            \textit{Scontain}
    \and
        Silvio Dragone\\
            \textit{IBM Research -- Zurich}
     \and
        Christof Fetzer\\
        \textit{TU Dresden}
}

\maketitle
\thispagestyle{plain}
\pagestyle{plain}

\errorstopmode


\begin{abstract}

High-assurance security systems require strong isolation from the untrusted world to protect the security-sensitive or privacy-sensitive data they process. 
Existing regulations impose that such systems must execute in a trustworthy operating system (OS) to ensure they are not collocated with untrusted software that might negatively impact their availability or security. However, the existing techniques to attest to the OS integrity fall short due to the cuckoo attack. In this paper, we first show a novel defense mechanism against the cuckoo attack, and we formally prove it. Then, we implement it as part of an integrity monitoring and enforcement framework that attests to the trustworthiness of the OS from $3.7\times$ to $8.5\times$ faster than the existing integrity monitoring systems. We demonstrate its practicality by protecting the execution of a real-world eHealth application, performing micro and macro-benchmarks, and assessing the security risk.

\glsresetall
\end{abstract}
\section{Introduction}
\label{sec:introduction}

\begin{figure}[tbp!]
    \centering
    \includegraphics[width=0.48\textwidth]{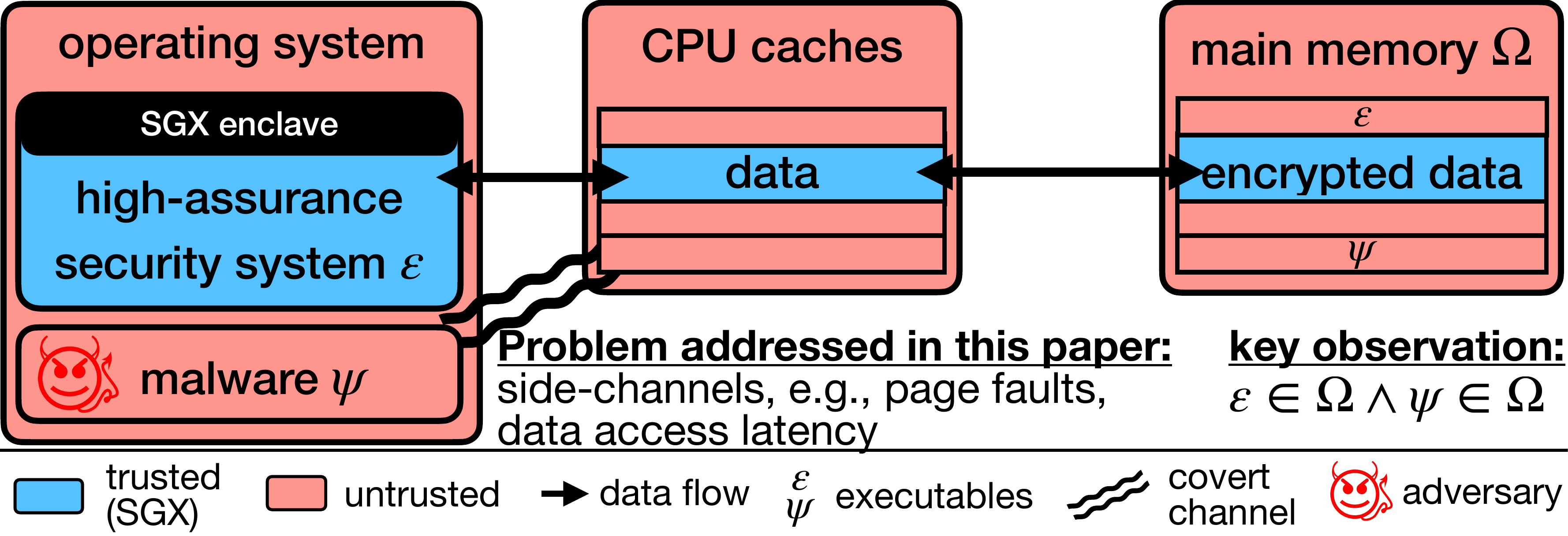}
    \caption{
        An adversary must run arbitrary software to mount a software side-channel attack that can compromise the confidentiality guarantee of Intel SGX.
        Colors are consistent across all figures.
   }
    \label{fig:SideChannel}
\end{figure}

High-assurance security systems~\cite{erezept, eperi2021top, fortranix_kms} leverage \glspl{tee}~\cite{costan2016intel, lee2020keystone, arm2009trustzone} because TEEs offer strong integrity and confidentiality guarantees in the face of untrusted privileged software, \ie, firmware, hypervisors, \gls{os}, and administrators. However, applications executing in a TEE cannot exist without the \gls{os}, which manages the computing resources and controls applications' life cycles. Thus, a \emph{trustworthy \gls{os}} is an essential element of each high-assurance security system because it guarantees its safety and security. Otherwise, an untrustworthy \gls{os} might run malware that halts the victim application or steals secrets from the TEE via side-channel attacks~\cite{Foreshadow, xu2015controlled}, as depicted in \autoref{fig:SideChannel}. Germany introduced regulations requiring high-assurance security systems in the eHealth domain~\cite{erezept} to execute inside \gls{tee} on a trustworthy \gls{os}~\cite{epa}. State-of-the-art mechanisms to attest to the OS's trustworthiness rely on the \gls{tpm}~\cite{tpm_2_0_spec}, a secure element storing and certifying integrity measurements of firmware and \gls{os}. Unfortunately, the TPM is vulnerable to the \emph{cuckoo attack} (a.k.a \emph{relay attack})~\cite{parno_bootstrapping, dhar2020proximitee} that makes the TPM attestation untrustworthy. We propose a novel defense mechanism against the TPM cuckoo attack, and we implement it as part of the framework responding to the German eHealth systems regulations~\cite{epa}.

\begin{figure}[bp!]
    \vspace{3mm}
    \centering
    \includegraphics[width=0.48\textwidth]{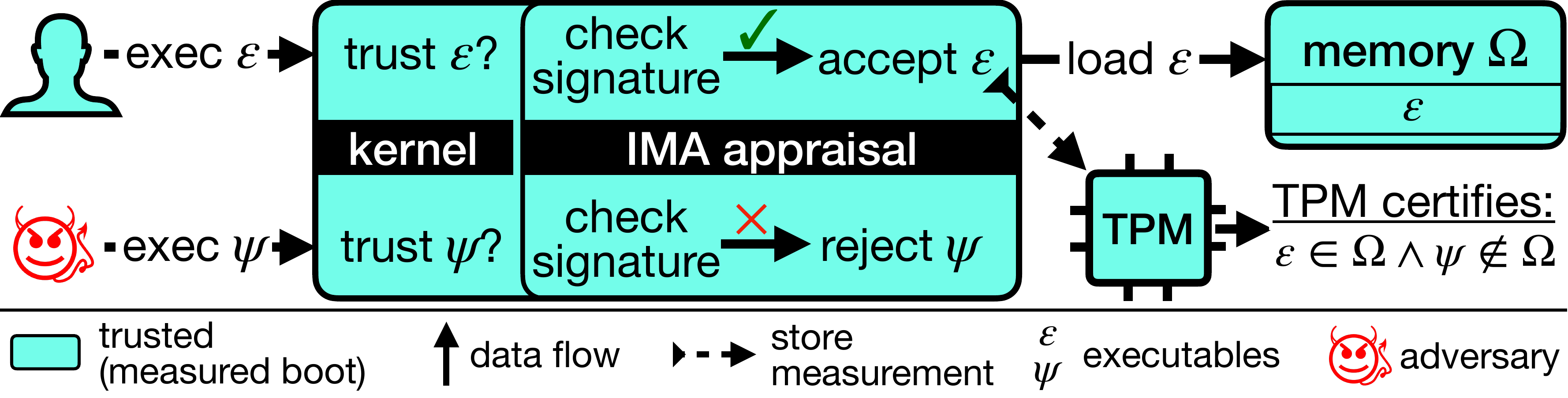}
    \caption{
        Integrity measurement architecture (IMA) is part of the kernel.
        It approves software to execute and provides reporting capacity to verify what software has been executed since the load of the kernel.
    }
    \label{fig:IMA}
\end{figure}

The \gls{ima}~\cite{tcg_ima_spec} and the \gls{drtm}~\cite{drtm_tcg} are state-of-the-art mechanisms providing OS integrity auditing and enforcement. The DRTM securely loads the kernel to the memory, and IMA, which is part of that kernel, ensures that the kernel loads only software whose integrity is certified with a digital signature. Both technologies, when used together, ensure the \emph{load-time integrity} of the kernel and software loaded to the memory during the OS runtime. Specifically, the DRTM, a hardware technology implemented in the CPU, stops all cores except one, disables interrupts, measures the to-be-loaded kernel, and executes the kernel with the IMA integrity enforcement mechanism. IMA restricts software loaded to the memory by reading the digital signature corresponding to the given software from the file system and verifying that this software's integrity measurement (a cryptographic hash over its binary) matches the original integrity measurement signed by a trusted party (\autoref{fig:IMA}). Thus, only software certified by a trusted party can be loaded to the memory by the kernel.

\begin{figure}[tbp!]
    \centering
    \includegraphics[width=0.48\textwidth]{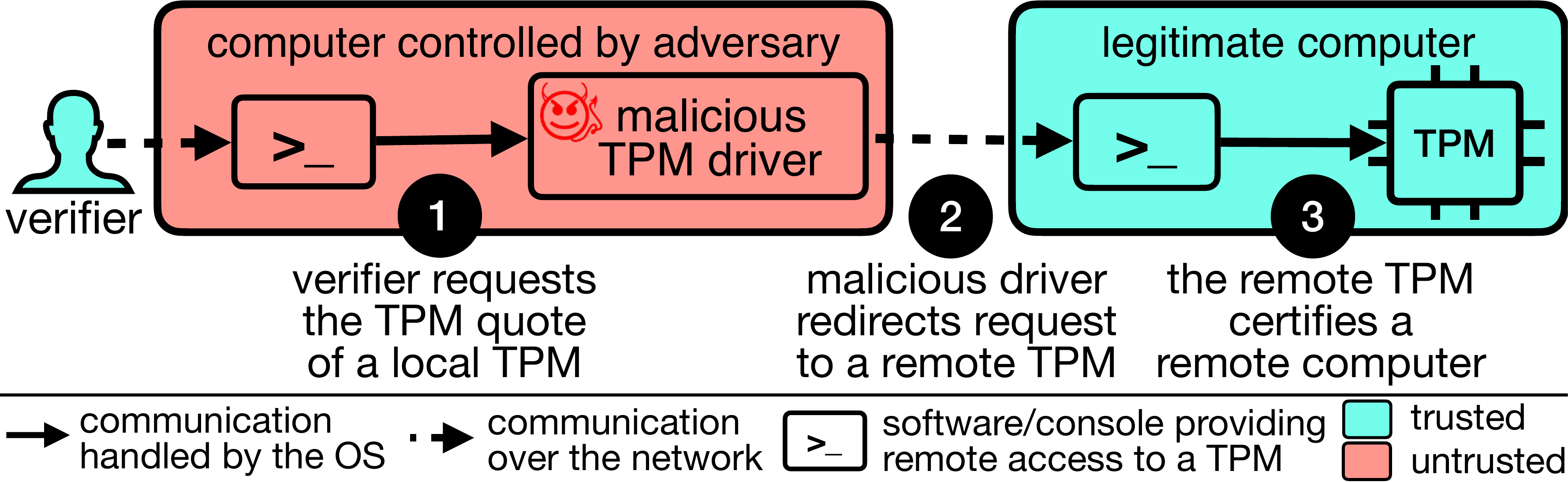}
    \caption{
        The \cuckooattack. The verifier connects to the compromised machine (left) and reads the TPM quote to verify its integrity.
        The \tpmquote is, however, retrieved from the remote TPM attached to a legitimate machine (right).
        The verifier cannot distinguish if the \tpmquote comes from the TPM attached to the local or remote machine.
    }
    \label{fig:Cuckoo}
\end{figure}

The TPM enables auditing of the kernel and software integrity because DRTM and IMA store corresponding integrity measurements in the tamper-proof TPM memory. The TPM then certifies the stored measurements to a verifier accordingly with the \emph{TPM remote attestation} protocol. However, the TPM remote attestation is prone to the cuckoo attack, which is a security issue for TPM-based systems~\cite{gligor2018establishing, 9095263, simtpm}. In this attack, an adversary certifies the software integrity of the underlying computer using certified measurements of another computer (see \autoref{fig:Cuckoo}). A \textit{verifier} connects to the compromised computer and communicates with the TPM to check the computer software integrity (\raisebox{-1pt}{\ding{202}}). The adversary prevents the verifier from accessing the local TPM by redirecting communication to a remote TPM (\raisebox{-1pt}{\ding{203}}). Consequently, the verifier reads the remote TPM, which attests to an arbitrary, trustworthy state (\raisebox{-1pt}{\ding{204}}), not the state of the compromised computer accessed by the verifier.

The existing defenses against the cuckoo attack have limited application in real-world \glspl{dc}. The first approach relies on the time side-channel~\cite{fink_catching_2011, seshadri2005pioneer} in which a remote TPM is unmasked by observing increased communication latency. This approach requires calculation of hardware-specific statistics, is prone to false positives because the high TPM communication latency (including signature generation) makes the distance bounding infeasible~\cite{parno_bootstrapping, 9095263}, and requires stable measurement conditions in which extraneous OS services are suspended during the TPM communication~\cite{fink_catching_2011} --- impractical assumptions for real-world \glspl{dc}. Flicker~\cite{flicker2008} adapts another approach. It exploits DRTM to run an application in isolation from the untrusted \gls{os}, allowing it to communicate with the TPM directly. Flicker is insufficient for the targeted systems like~\cite{erezept} because i) it does not attest to the computer location, making the DRTM attestation untrustworthy because of simple hardware attacks~\cite{winter2013hijacker} and cold-boot attacks~\cite{halderman2009lest} and ii) while it permits to split applications in multiple services that run isolated, it does not support systems with moderate throughput and latency requirements. In more detail, DRTM provides isolation in which the entire CPU executes \emph{only a single service at a time} and a single context-switching takes 10-100s of milliseconds~\cite{flicker2008, mccune_trustvisor:_2010}. It results in an estimated program execution's throughput of about 1-10 requests per computer per second when running multiple eHealth services, like \cite{epa}. A practical solution requires that hundreds of services are processed in parallel per computer. We require an improvement of at least one order of magnitude in throughput compared to Flicker. Other approaches~\cite{danisevskis2015graphical, nunes2021root} fall short in the context of the TPM because i) the TPM is a passive device controlled by software that could counterfeit its communication with external devices and ii) they would require human interaction during each computer boot. 

The limitations of the existing solutions motivate us to propose a new automatic, practical at the data center-scale defense mechanism that deterministically detects the cuckoo attack and allows for the processing of parallel requests. We demonstrate that despite the differences in their threat models and designs, \gls{tee} and TPM-based techniques complement each other, allowing for mitigating the cuckoo attack. Consequently, high-assurance security systems executing inside TEE can attest to the OS integrity. Our solution builds trust in a remote computer starting from a piece of code executing inside the TEE, and then systematically extend it to the entire OS. First, we leverage TEE to settle a trusted piece of code on an untrusted remote computer. We use it to verify that the computer is in the correct \gls{dc} and mitigate the cuckoo attack. This allows us to extend trust to the TPM, then to the loaded kernel and its integrity-enforcement mechanism and, finally, to software being executed during the OS runtime. 

We implement this approach in an integrity monitoring and enforcement framework called \sys, which ensures that high-assurance security applications execute on correctly initialized and integrity-enforced OS located in the expected \gls{dc}. The high-assurance security systems conform to the TEE threat model, while they gain OS integrity guarantees under a less rigorous threat model typical for TPM-based systems. We perform security risk analysis related to the use of these techniques in \S\ref{eval:risk}.

Altogether, we make the following contributions:
\begin{enumerate} 
	\item {We designed and implemented an integrity monitoring and enforcement framework called \sys that: }
		\begin{itemize}
			\item attests to the OS trustworthiness (\S\ref{sec:introduction},\S\ref{sec:background}),
			\item defends against the cuckoo attack (\S\ref{sec:initialization}, \S\ref{impl:establish_trust}),
			\item provides a reliable approach to estimate the geolocation of physical servers beyond the simple TPM geo-tagging (\S\ref{over:trusted_beacons}),
			\item provides local attestation, allowing decentralization of the monitoring system (\S\ref{sec:highlevelarch}, \S\ref{over:verification_protocol}), 
			\item the service itself can be remotely attested (\S\ref{impl:policy_verification}),
			\item verifies the compliance of provisioned resources with a given policy (\S\ref{sec:policy}, \S\ref{over:verification_protocol}).			
		\end{itemize} 
	\item We assessed the security risk of \sys (\S\ref{eval:risk}).
	\item We demonstrated \sys protecting a real-world application in the eHealth domain (\S\ref{eval:realworld}).
	\item We evaluated its security and performance (\S\ref{sec:evaluation}).
	\item We provided the formal proof of the protocol detecting the \cuckooattack (\S\ref{sec:evaluation:formal_analysis}).
\end{enumerate}

\section{Threat Model}
\label{sec:threat_model}

We adopt the threat model of organizations, such as governments, banks, and health, legally bound to protect the security-sensitive data they process. In particular, we assume they execute high-assurance security systems in their own \glspl{dc} or in the hybrid cloud in which security-critical resources are provisioned on-premises. This implies limited and well-controlled access to \glspl{dc}, allowing us to assume that an adversary, \eg, a rogue operator, cannot perform physical or hardware attacks. To ensure that a high-assurance security system executes inside the \gls{dc}, we only presume that dedicated computers, called \emph{trusted beacons}, are located inside that \gls{dc} and cannot be physically moved outside (\S\ref{over:trusted_beacons}).

Initially, we only trust the CPU (including its hardware features TEE and DRTM) and a small piece of code (the \emph{agent}). Using the TEE attestation protocol, we ensure that the legitimate agent executes inside the TEE on a genuine CPU on some computer. Then, we use the agent to verify that the computer is located in the correct DC by measuring the proximity to the trusted beacon via a round-trip time distance-bounding protocol. Once we ensure that the agent runs in the expected DC (no physical and hardware attacks), we use it to establish trust with the local TPM with the help of our protocol formally proved to be resistant to the cuckoo attack (\S\ref{sec:evaluation:formal_analysis}). At this point, we use the TPM to extend the trust to the kernel and its built-in integrity-enforcement mechanism, IMA. Eventually, we use IMA to expand trust to the software loaded during the OS runtime. 

High-assurance security systems executing inside the TEE follow the TEE threat model, \ie, operating system, firmware, other software, and system administrator are untrusted. The additional guarantees of the operating system integrity follow the threat model of TPM-based systems, \ie, software whose integrity is enforced at load-time behaves in a trustworthy way also during its execution. The runtime integrity of the process can be enforced using existing techniques, such as control-flow integrity enforcement~\cite{mustakimur2019origin}, fuzzing~\cite{fuzzing}, formal proofs~\cite{hacl}, memory-safe languages~\cite{matsakis_rust_2014}, or memory corruption mitigation techniques (position-independent executables, stack-smashing protection, relocation read-only techniques). Please note that many of these techniques are applied nowadays by default during the software packaging process, as in the case of Alpine Linux~\cite{alpine_linux}.

We assume a financially or governmentally motivated adversary who might gain root access to selected computers inside a \gls{dc} by exploiting network or OS misconfigurations, exploiting vulnerabilities in the \gls{os}, or using social engineering. Her goal is to extract security-sensitive or privacy-sensitive data, \eg, personal data, credentials, or cryptographic material. She can stop or halt individual computers or processes, but she cannot stop all central monitoring service instances responsible for reporting security incidents. We consider an untrusted network where an adversary can view, inject, drop, and alter messages. She can call the API with any parameters and configure the routing, forcing packages to choose faster or slower routes. Our network model is consistent with the classic Dolev-Yao adversary model~\cite{DolevYao}. We rely on the soundness of the employed cryptographic primitives used within software and hardware components.
\section{Design Decisions}
\label{sec:background}

Our objective is to provide a design that:
i) enforces that only trusted software is executed on a computer;
ii) monitors the remote computer \gls{os} to verify compliance to integrity requirements;
iii) allows high-assurance security systems to get insights into the \glspl{os} integrity.

\begin{figure}[tbp!]
    \centering
    \includegraphics[width=0.48\textwidth]{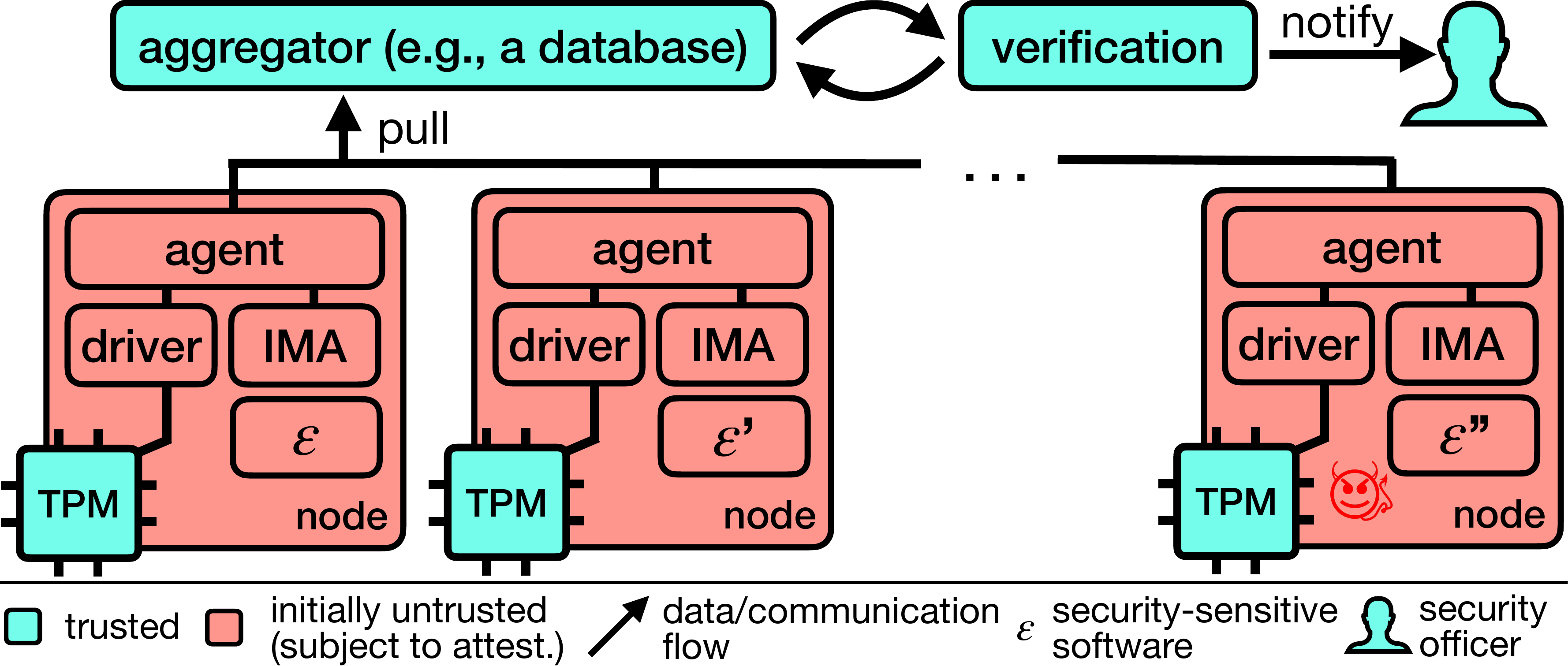}
    \caption{
        The architecture of existing integrity monitoring systems. 
        The security officer uses a monitoring system to verify that high-assurance security systems execute on hosts running trusted software.
    }
    \label{fig:monitoring_architecture}
\end{figure}

We start by introducing the existing integrity monitoring systems architecture \cite{intel_secl, ibm_tpm_acs, opencit_01_org} and adjust it to meet the security guarantees required by high-assurance security systems. \autoref{fig:monitoring_architecture} shows the integrity monitoring architecture where a central server pulls integrity measurements from computers by communicating with dedicated software, \emph{the agent}.
The agent on each computer collects data from the underlying security and auditing subsystems that measure and enforce the OS integrity.
Central servers aggregate the data in databases, verify it against whitelists, and notify the \emph{security officer} about integrity violations.
Such architecture relies on the TPM as a root of trust.

\begin{shaded*}
	\begin{enumerate}[wide = 0pt, nosep,leftmargin=!,font=\bfseries,start=1]
		\item Enforce the load-time integrity with secure boot and OS integrity enforcement.
	\end{enumerate}
\end{shaded*}

\emph{Secure boot} \cite{wilkins_secure_boot_2013} is the state-of-the-art technology to enforce that only trusted software bootstraps a computer.
It relies on the \emph{chain of trust} where each component measures the integrity (calculates a cryptographic hash) of the next component and executes it only if the hash matches a corresponding digital signature.
The \emph{measured boot}~\cite{tcg_srtm_bios_spec, tcg_srtm_uefi_spec} complements it by storing hashes in the TPM, thus enabling auditing.

\label{sec:LinuxIMA}
The \acrfull{ima}~\cite{ima_design_2004, tcg_ima_spec} extends the functionality of measured boot and secure boot to the OS level.
IMA is part of the kernel and verifies all files' integrity (\ie, executables, configuration files, dynamic libraries) before they are loaded to the memory.
In particular, IMA-appraisal~\cite{ima_appraisal} enforces that the kernel loads files whose hashes are certified with digital signatures stored in the file system (\autoref{fig:IMA}).
The application execution is halted until a dynamic library is loaded, and fails if the library fails the integrity check.
IMA enables auditing by maintaining an \emph{\imalog}, a dedicated file storing hashes of all files loaded to the memory since the kernel load.
It adds each file to the \imalog and stores a hash over it in the TPM before the file is loaded to the memory.
Any tampering of the \imalog is detectable because the \imalog's integrity hash must match the value stored in the TPM.

\begin{shaded*}
	\begin{enumerate}[wide = 0pt, nosep,leftmargin=!,font=\bfseries,start=2]
		\item Enable remote attestation to prove that secure boot and integrity enforcement are enabled.
	\end{enumerate}
\end{shaded*} 

The TPM remote attestation protocol~\cite{tcg_tpm_attestation} delivers a technical assurance of the computer's integrity. The TPM chip digitally signs a report (\emph{\tpmquote}) certifying hashes recorded since the computer boot. The hashes reflect loaded firmware and kernel and prove that integrity enforcement mechanisms are enabled. The verifier can check that the \tpmquote has not been manipulated because the TPM signs the \tpmquote with a signing key that is embedded in the TPM and linked to the \gls{ca} of the TPM manufacturer. However, the monitoring system cannot merely rely on the \gls{tpm} attestation because it is vulnerable to the \cuckooattack~\cite{parno_bootstrapping}. It is indistinguishable whether an untrusted OS proves its integrity presenting a \tpmquote from a local TPM or impersonates a trustworthy OS presenting a \tpmquote from a remote TPM.

\begin{shaded*}
	\begin{enumerate}[wide = 0pt, nosep,leftmargin=!,font=\bfseries,start=3]
		\item Detect the \cuckooattack by authenticating the TPM with a secret random number.
	\end{enumerate}
\end{shaded*}

\begin{figure}[tbp!]
    \centering
    \includegraphics[width=0.48\textwidth]{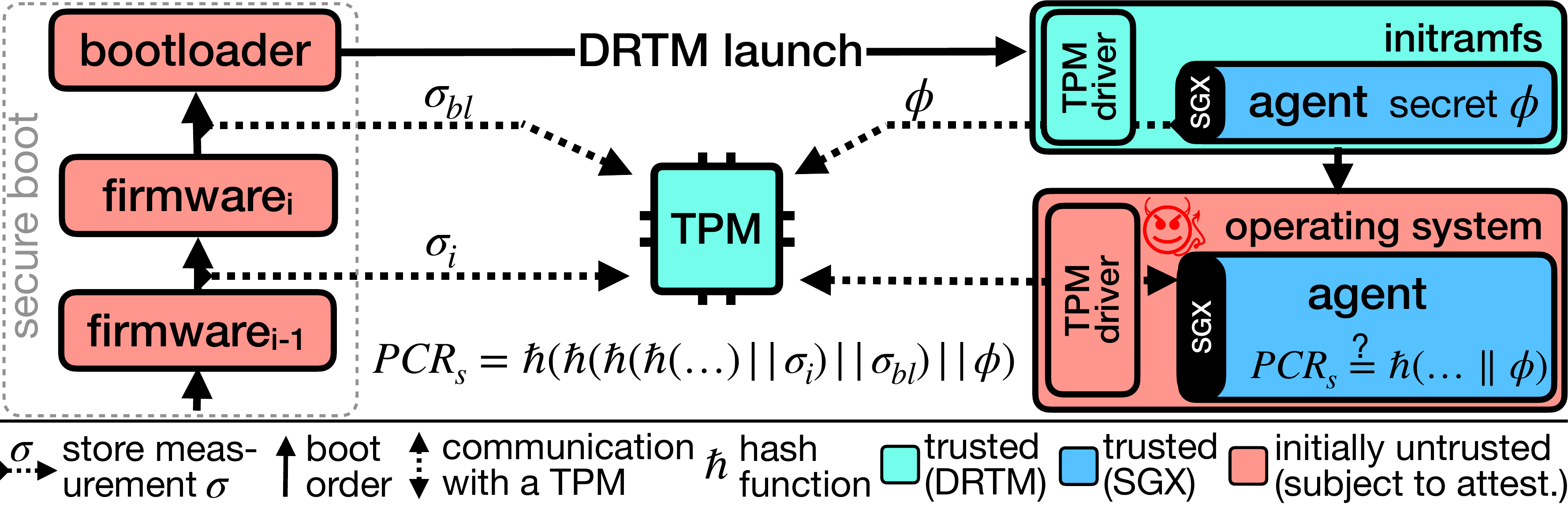}
    \caption{
        Defense against the cuckoo attack.
        The agent shares with the TPM a randomly generated secret $\phi$, which is used later to authenticate the TPM.
        PCR is the TPM tamper-resistant memory.
    }
    \label{fig:secret}
\end{figure}

The monitoring system must ensure that the \tpmquote originated from the local TPM, \ie, the TPM that collected integrity measurements from the software components that booted the \gls{os} on the underlying computer.
We propose to extend the agent with the functionality of checking that it communicates with the local TPM.
The general idea consists of sharing a randomly generated \emph{secret $\phi$} with the local TPM to identify it uniquely and then use the secret to authenticate the TPM (\autoref{fig:secret}).
The main challenge is generating a secret and sharing it with the local TPM without revealing it to an adversary.
The main challenge is how to generate a secret and share it with the local TPM without revealing it to the adversary.
Otherwise, the adversary can mount the \cuckooattack by sharing it with a remote TPM.

\begin{shaded*}
	\begin{enumerate}[wide=0pt, nosep,leftmargin=!,font=\bfseries,start=4]
		\item Protect the secret in the TPM by relying on the one-way cryptographic hash function.
	\end{enumerate}
\end{shaded*}

The TPM contains dedicated memory registers, called \glspl{pcr}, that have important properties; they cannot be written directly, but they can only be extended with a new value using a cryptographic one-way hash function.
The operation can be expressed as: {\emph{PCR\_extend(n,value): pcr[n] = hash(pcr[n]||value)}}. 
We propose to extend the secret $\phi$ on top of the existing measurements stored in the PCR to achieve the following properties:
i) an adversary cannot extract the secret from the PCR value after the secret is extended to the PCR because the hash function result is not invertible;
ii) an adversary cannot reproduce the PCR value in another TPM without knowing the secret, or finding a collision in the hash function;
iii) after extending the TPM with the secret, the secret is no longer needed to identify the TPM because the PCR value extended with the secret is unique.

\begin{shaded*}
	\begin{enumerate}[wide=0pt, nosep,leftmargin=!,font=\bfseries,start=5]
		\item Leverage DRTM technology to provide a trusted and measured environment to access the local TPM.
	\end{enumerate}
\end{shaded*}

We must ensure that the secret is shared with the local TPM securely. We do it in a trusted environment established by hardware technologies available in modern CPUs because these technologies also permit verification of the established execution environment's integrity. 
Therefore, they allow detecting (post-factum) any secret extraction attempt, including software side-channel attacks, because such attacks require violating the kernel or initramfs integrity.

We propose generating the secret and extending it to PCRs inside the initramfs~\footnote{The initramfs is a minimalistic root filesystem that provides a user space to perform initialization tasks, like loading device drivers, mounting network file systems, or decrypting a filesystem \cite{rootfs_encryption_2005}, before the OS is loaded.} because DRTM allows for later verification of the kernel and initramfs integrity. Specifically, the \gls{drtm}~\cite{drtm_tcg}, which is a hardware technology that establishes an isolated execution environment to run code on a potentially untrusted computer, can be used during the boot process (\ie, by tboot~\cite{tboot}) to provide a measured load of the Linux kernel and initramfs. 

The integrity measurements performed by DRTM cannot be forged because the TPM offers a dedicated range of PCRs (\emph{dynamic PCRs}) that can only be reset or extended when the TPM is in a certain \emph{locality}~\cite{JayaramMasti_2013}; Only the code executed by DRTM can enter such locality. Therefore, the presence of measurements in dynamic PCRs confirms that the DRTM was executed, and the comparison of PCRs with the golden values confirms that the secret was shared with the local TPM because the correct TPM driver was used. 

\begin{shaded*}
	\begin{enumerate}[wide=0pt, nosep,leftmargin=!,font=\bfseries,start=6]
		\item Leverage Intel SGX to transfer the golden TPM PCR value to the OS runtime securely.
	\end{enumerate}
\end{shaded*}

Once the secret is shared with the TPM, we must expose the unique local TPM's identifier (PCR value extended with the secret) to the agent running in the OS.
To do so, we leverage \gls{sgx}~\cite{costan2016intel}, a hardware CPU extension that provides confidentiality and integrity guarantees to the code executed in so-called \emph{enclaves} in the presence of an adversary with root access to the computer. 
It offers a \emph{sealing}~\cite{anati2013innovative} property that permits storing a secret on an untrusted disk where only the same enclave running on the same CPU can read it. 
The sealing and its revert operation \emph{unsealing} use a CPU- and an enclave-specific key to encrypt and sign data in untrusted storage. 
We propose to communicate with the TPM from the inside of an enclave. First, the enclave executes in the initramfs where it shares a secret with the local TPM and seals the expected value of the TPM PCR to the disk. 
Then, it executes in the untrusted OS, where it authenticates the TPM using the PCR value unsealed from the disk.

\begin{shaded*}
	\begin{enumerate}[wide=0pt, nosep,leftmargin=!,font=\bfseries,start=7]
		\item Leverage the SGX local and remote attestation to expose integrity measurements to the verifiers.
	\end{enumerate}
\end{shaded*}

\gls{sgx} offers local and remote attestation protocols \cite{johnson2016intel}.
While both protocols allow verifying that the expected code runs on a genuine Intel CPU, the SGX local attestation also permits two enclaves to learn that they execute on the same CPU.
We rely on this property to permit high-assurance security systems to establish trust with the agent running on the same computer.
Like this, high-assurance security systems gain access to integrity measurements of the surrounding OS. 
Similarly, central monitoring services leverage the SGX remote attestation to establish trust with agents.

\begin{shaded*}
 	\begin{enumerate}[wide=0pt, nosep,leftmargin=!,font=\bfseries,start=8]
 		\item Formally prove the protocol of establishing trust between the agent and the TPM.
 	\end{enumerate}
\end{shaded*}

We use formal verification techniques to prove that the \sys protocol is resilient against the cuckoo attack because functional software testing cannot detect protocol errors since they only appear in the presence of a malicious adversary. We rely on automated security protocol verification approaches~\cite{SAPIC, proVerif, AVISPA} because they can provide guarantees of the protocol's correctness~\cite{FRWexample1,FRWexample2,FRWexample3}. Specifically, we use SAPIC~\cite{SAPIC} tool to implement a formal model of the \sys protocol, verify its integrity, and prove that it is resilient against the cuckoo attack (\S\ref{sec:evaluation:formal_analysis}).


\section{\sys architecture}
\label{sec:overview}

\begin{figure}[tbp!]
  \centering
  \includegraphics[width=0.48\textwidth]{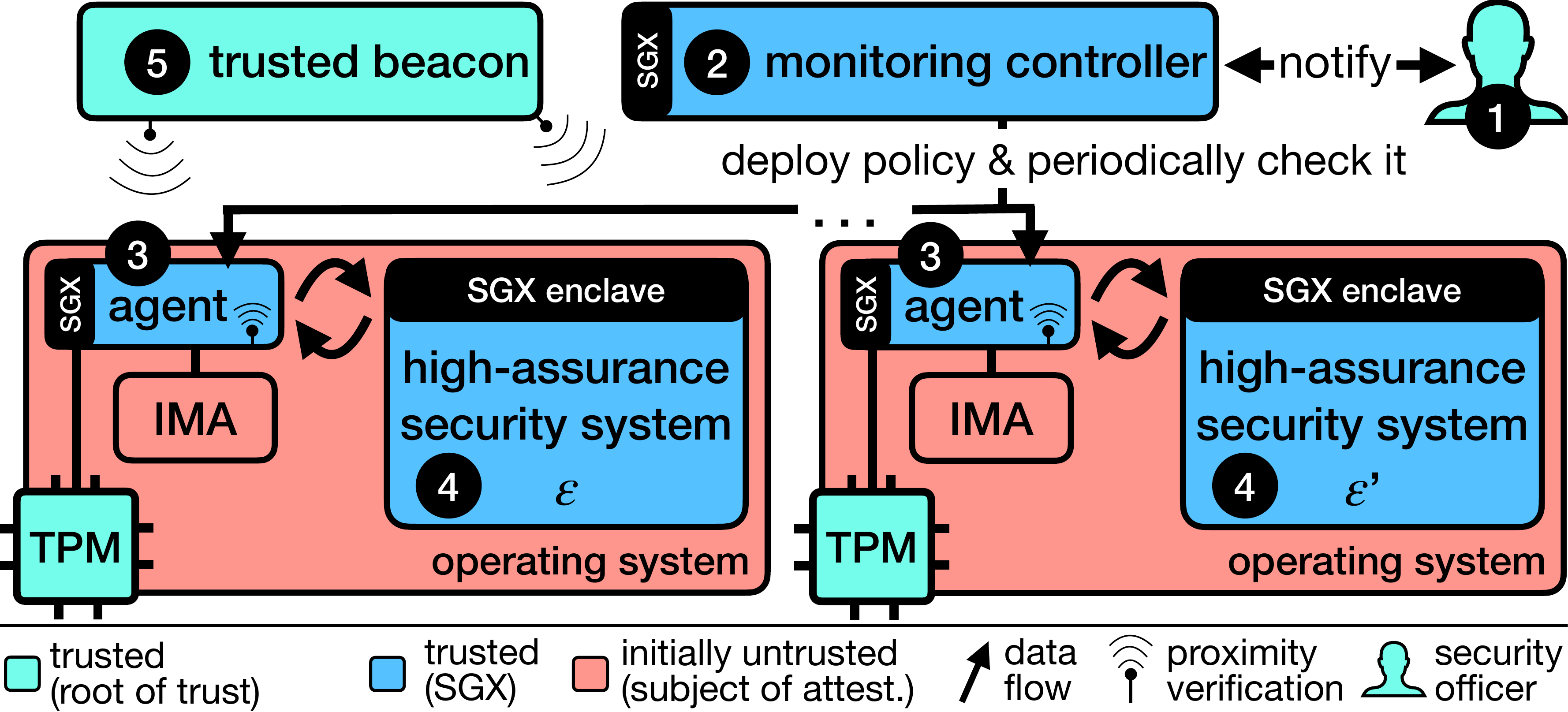}
  \caption{
      \sys architecture. 
      The agent provides access to integrity measurements certified by the local TPM after mitigating the cuckoo attack. 
      High-assurance security system $\varepsilon$ and the monitoring controller query the agent to verify the computer geolocation and operating system integrity.
  }
  \label{fig:SystemArchitecture}
\end{figure}

\subsection{High-level Overview}
\label{sec:highlevelarch}

\autoref{fig:SystemArchitecture} shows a high-level overview of the \sys architecture, which consists of five entities.
A security officer (\raisebox{-1pt}{\ding{202}}) uses a controller (\raisebox{-1pt}{\ding{203}}) to define security policies describing correct (trusted) OS configurations. 
The controller communicates with agents (\raisebox{-1pt}{\ding{204}}) running on every computer to check whether high-assurance security systems (\raisebox{-1pt}{\ding{205}}) are executed in a trusted environment defined in security policies.
Both the controller (\raisebox{-1pt}{\ding{203}}) and the high-assurance security system executing inside SGX (\raisebox{-1pt}{\ding{205}}) systematically query the agent to check if the operating system integrity conforms to the criteria defined inside a security policy.
Note that the integrity measurements are not aggregated or verified centrally.
Instead, agents aggregate them and verify them locally on computers.
Agents verify their location using trusted beacons (\raisebox{-1pt}{\ding{206}}), services running in a known geographical location, \ie, specific \glspl{dc}.

We distinguish between two types of verifiers communicating with agents, local and remote verifiers.
A local verifier is a high-assurance security system that requires strong confidentiality guarantees (\raisebox{-1pt}{\ding{205}}). 
An example of such a service is a key management system~\cite{chakrabarti2017intel, fortranix_kms, palaemon_2020} that executes inside an \gls{sgx} enclave to protect integrity and confidentiality against privileged adversaries.
The local verifier detects violations of the operating system integrity by communicating with the agent running on the same host.

A remote verifier, \eg, (\raisebox{-1pt}{\ding{203}}), is an application running on a different computer than the agent.
It aims to verify that the remote computer is located in the specific \gls{dc} and its OS is in the expected state.
Typically, a remote verifier checks the integrity of the distributed system's deployment, \ie, various services distributed over machines, data centers, and availability zones.
The controller has broader knowledge about the network load, machine failures, service migrations, software updates. 
It helps the security officer to manage the deployment while relying on individual services to react autonomously to integrity violations.
The controller might be part of the \gls{siem} system that correlates system behavior to detect multi-faceted attacks~\cite{bhatt2014siem}.

\subsection{Policy}
\label{sec:policy}

\lstdefinestyle{interfaces}{
  float=tbp,
  floatplacement=tbp,
  abovecaptionskip=-20pt,
  xleftmargin=0.3cm
}
\begin{lstlisting}[caption=Policy example,label={lst:policy},language=yaml,breaklines=true,breakatwhitespace=true,postbreak=\mbox{\textcolor{red}{$\Rdsh$}\space},style=interfaces,escapechar=^]
chain: |-  ^\label{policy_chain}^
  -----BEGIN CERTIFICATE-----
  # TPM manufacturer certificates
  -----END CERTIFICATE-----
whitelist: 
  - pcrs:
# secure boot / measured boot, PCRs 0-9
      - {id: 0, sha256: ff0c...e3} ^\label{policy_staticPcrsStart}^      
      - {id: 3, sha256: e850...3e} ^\label{policy_staticPcrsEnd}^
# trusted boot (DRTM) PCRs, 17-19
      - {id: 18, sha256: f9d0...cb} ^\label{policy_dynamicPcrsStart}^
      - {id: 19, sha256: a1e7...00} ^\label{policy_dynamicPcrsEnd}^
runtime:
  certificate: |- ^\label{policy:softwareCertificate}^
      -----BEGIN CERTIFICATE-----
      # IMA uses a certificate to verify signatures
      -----END CERTIFICATE-----
  software: ^\label{policy:softwareWhitelistStart}^  
      - name: agent-0.8.0
        whitelist:
          840f...72: /bin/agent    
      - name: AppArmour ^\label{policy:sshdStart}^                         
        whitelist:
# hash of the executable        
          1e73...f6: /sbin/apparmour
# hash of the configuration file
          c39e...34:  /etc/apparmour^\label{policy:sshdEnd}^ ^\label{policy:softwareWhitelistEnd}^                        
location:                     
  - host: https://datacenter:10000/beacon ^\label{policy:beaconsStart}^ 
    max_latency: 2 # in milliseconds
    chain: |- 
      -----BEGIN CERTIFICATE-----
      # TLS certificate chain of the trusted beacon
      -----END CERTIFICATE----- ^\label{policy:beaconsEnd}^    
\end{lstlisting}

The security officer defines security policies (\eg \autoref{lst:policy}) to declaratively state what software and dynamic libraries are permitted to run on the computer and what is the proper OS configuration.
He creates distinct security policies for each high-assurance security system. For example, a key management system has a different policy than a system processing medical data because they use different dynamic libraries, software, and OS configurations.
The monitoring controller reduces the burden of creating policies by allowing defining templates that can be combined to build individual policies with overlapping configurations.
For example, services running on the same type of OS share the same template that describes software and configuration specific to that OS. 

The agent uses the security policy to verify the OS integrity. The OS is trusted if and only if the load-time integrity measurements of the kernel and the load-time integrity measurements of files loaded to the memory during the OS runtime are declared on the whitelist or their corresponding digital signatures are verifiable using the certificate declared in the policy. 

In more detail, the agent uses the TPM manufacturer's CA certificate chain to verify that the TPM chip attached to the computer is legitimate (line \ref{policy_chain}). The integrity of firmware and its configuration is represented as a whitelist of static PCRs (lines \ref{policy_staticPcrsStart}-\ref{policy_staticPcrsEnd}), while the integrity of the Linux kernel and the initramfs is specified as a whitelist of dynamic PCRs (lines \ref{policy_dynamicPcrsStart}-\ref{policy_dynamicPcrsEnd}). Trusted configuration files, executables, and dynamic libraries are defined in the form of hashes (lines \ref{policy:softwareWhitelistStart}-\ref{policy:softwareWhitelistEnd}) and a signing certificate (line \ref{policy:softwareCertificate}). Software updates are supported via complementary solutions~\cite{tsr_2020, imasig_updates} and require specification of the certificate in the policy (line \ref{policy:softwareCertificate}).

\begin{figure}[tbp!]
  \centering
  \includegraphics[width=0.48\textwidth]{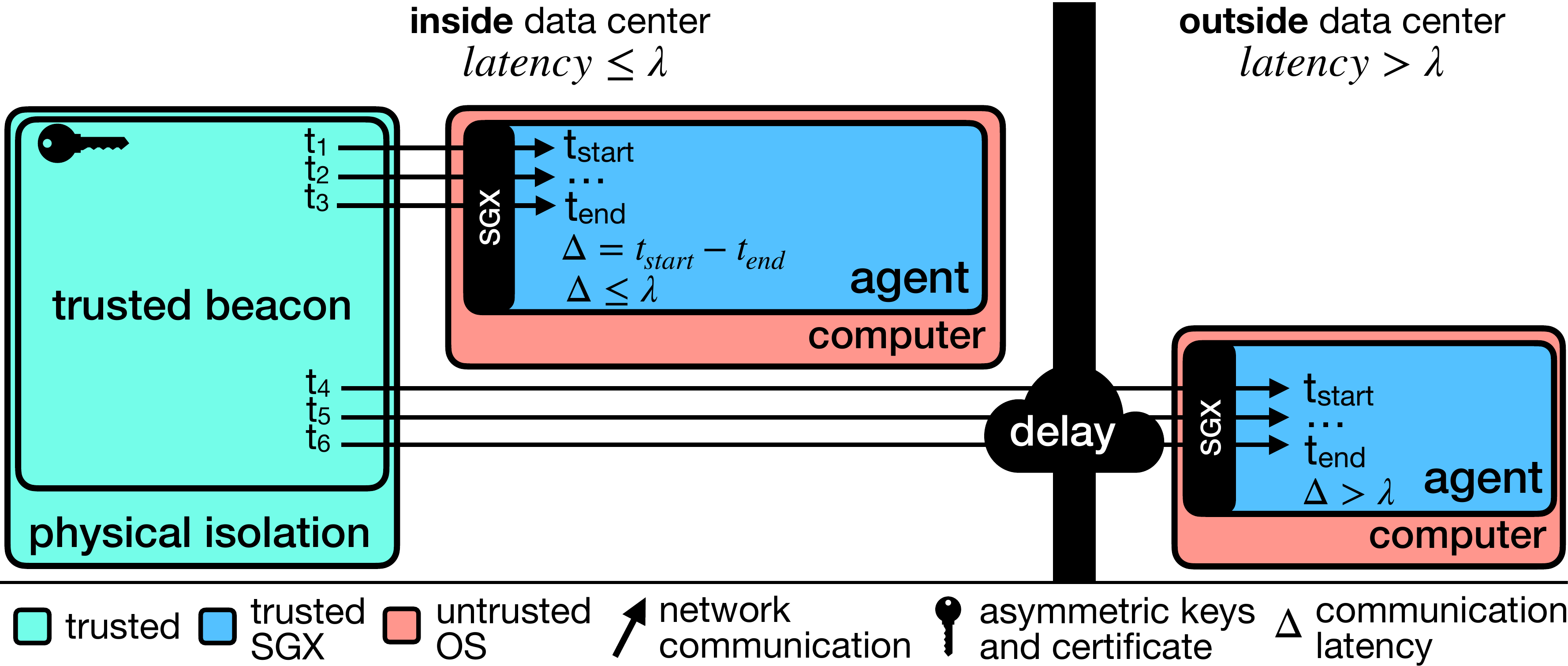}
  \caption{
      Trusted beacons. 
      Agents rely on the trusted beacon to check that they are located in the expected data center.
      Only machines located inside the same data centers can achieve very low network latency required to prove their proximity.
  }
  \label{fig:trusted_beacon}
\end{figure}

\subsection{Trusted Beacon}
\label{over:trusted_beacons}
A policy might constrain the computers' proximity to the well-known trusted beacons deployed in \glspl{dc} (lines~\ref{policy:beaconsStart}-\ref{policy:beaconsEnd}).
A trusted beacon is a network service that responds to agents' requests with the current timestamp. The agent can then estimate the physical machine's proximity by measuring the network communication's round-trip times. The adversary cannot accelerate network packets enough to achieve a very short round-trip time achievable only between machines in the same local network.

\autoref{fig:trusted_beacon} shows a high-level view of the trusted beacon proximity verification protocol. The trusted beacon contains the asymmetric keypair with a certificate issued by a trusted authority, \eg, a \gls{dc} owner. These credentials, known only to the trusted beacon, prove that the \gls{dc} owner placed the trusted beacon in the \gls{dc}, and the trusted beacon executes in a trusted environment. 
The agent establishes trust with the trusted beacon by reading timestamps signed by the trusted beacon. The agent then estimates the network latency by calculating a trimmed mean from the differences between timestamps obtained from pairs of consecutive requests. A trimmed mean allows for tolerating network latency fluctuations because it excludes outliers.

Our design does not restrict what security mechanisms must protect the trusted beacon. In particular, the trusted beacon could be a network-accessible hardware security module (HSM)~\cite{ibmCryptoCard4769} returning signed timestamps. HSM is a crypto coprocessor offering the highest level of security against software and hardware attacks. It is embedded in a tamper responsive enclosure to actively detect physical and hardware attacks and protect against side-channel attacks. A cheaper but less secure alternative might run a TEE-based application implementing the abovementioned protocol over TLS. Related work~\cite{dhar2020proximitee} demonstrated that the network communication round-trip time between two SGX enclaves located in the same network take in average $264\,\mu$s, a latency not achievable from the outside of the data center.

\subsection{Policy Verification Protocol}
\label{over:verification_protocol}

\begin{figure}[tbp!]
  \centering
  \includegraphics[width=0.48\textwidth]{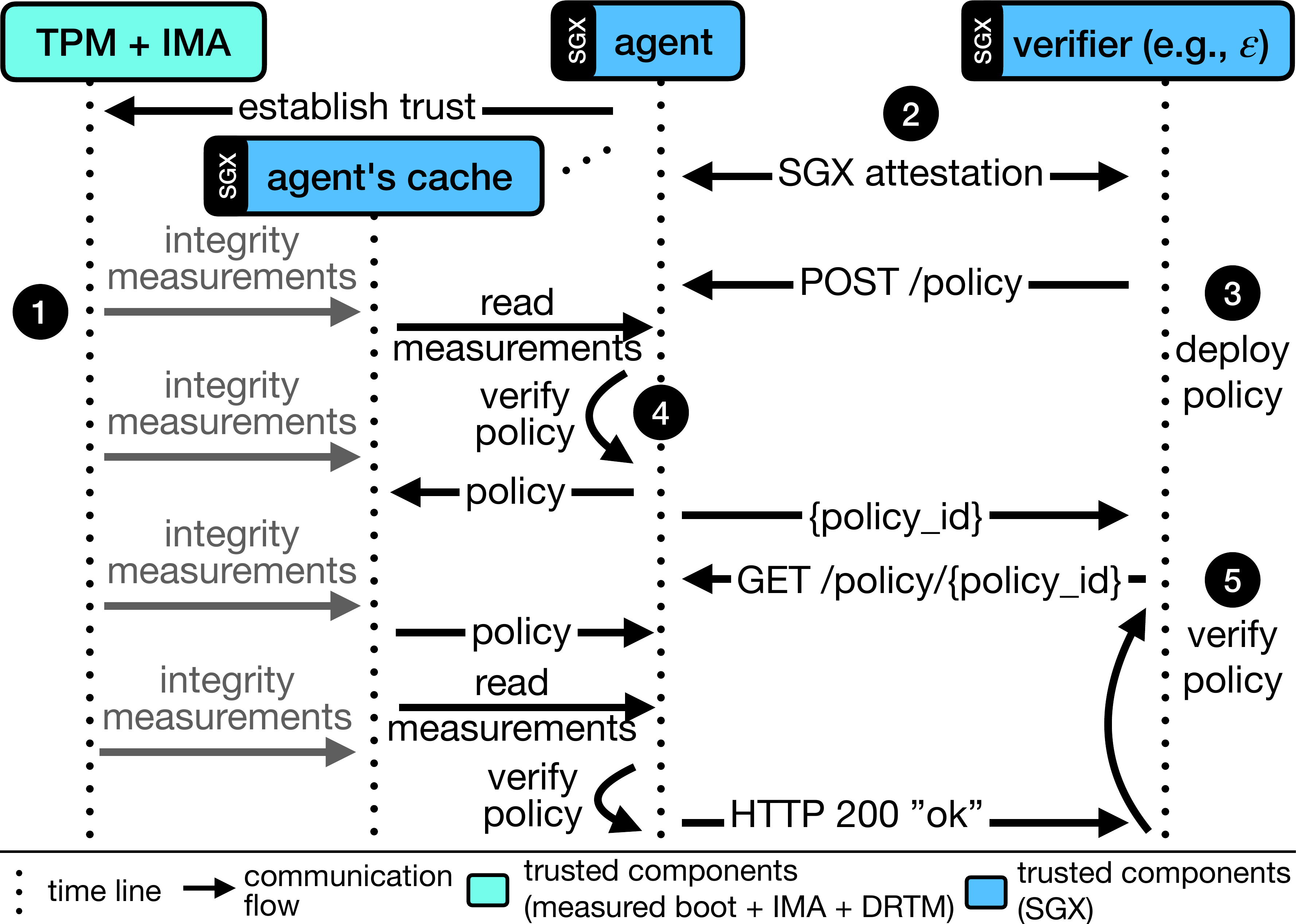}
  \caption{
      \sys policy verification protocol. 
      The agent maintains a separate thread (agent's cache) to constantly read the platform's fresh integrity measurements. 
      Verifiers query the agent in parallel to ensure the compliance of the platform to the policy.
  }
  \label{fig:PolicyEnforcement}
\end{figure}

We designed the agent to act as a facade between the verifier and the TPM to enable multiple verifiers to check the OS integrity concurrently.
\autoref{fig:PolicyEnforcement} shows how a verifier uses the policy verification protocol to attest to the OS integrity. 
The agent regularly reads the list of new software loaded by the \gls{os}, the \tpmquote, and persists it into the \emph{cache} that reduces the policy verification latency for future requests (\raisebox{-1pt}{\ding{202}}). 
The local or remote verifier perform the SGX local or remote attestation~\cite{johnson2016intel} to verify the agent's identity and integrity and the CPU genuineness. 
The local attestation also proves that the agent runs on the same CPU (\raisebox{-1pt}{\ding{203}}).
Once the verifier deploys the policy (\raisebox{-1pt}{\ding{204}}), the agent checks that the computer complies with the policy, stores the policy, and returns the corresponding \emph{policy\_id} (\raisebox{-1pt}{\ding{205}}). 
The verifier uses the \emph{policy\_id} to re-evaluate the policy during future health checks (\raisebox{-1pt}{\ding{206}}).

\section{Implementation}
\label{sec:implementation}
We implemented \sys on top of the Linux kernel.
We use existing integrity enforcement mechanisms built in the Linux kernel, \ie, IMA-appraisal, kernel module signature verification, and AppArmor.
We rely on the support for the secure boot built-in the underlying firmware.
We developed remote attestation components, \ie, the agent in memory-safe language Rust~\cite{matsakis_rust_2014} and the monitoring controller in Python. 
We implemented the cuckoo attack detection mechanism and the policy verification protocol inside the agent. 
The monitoring controller allows defining policies, verifying the remote computer system's integrity, and alerting about integrity violations.
We rely on the SCONE framework \cite{arnautov2016scone} and the SCONE cross-compiler to run \sys inside the \gls{sgx} enclave. 

\subsection{Computer bootstrap}
\label{sec:initialization}

\begin{figure}[tbp!]
    \centering
    \includegraphics[width=0.48\textwidth]{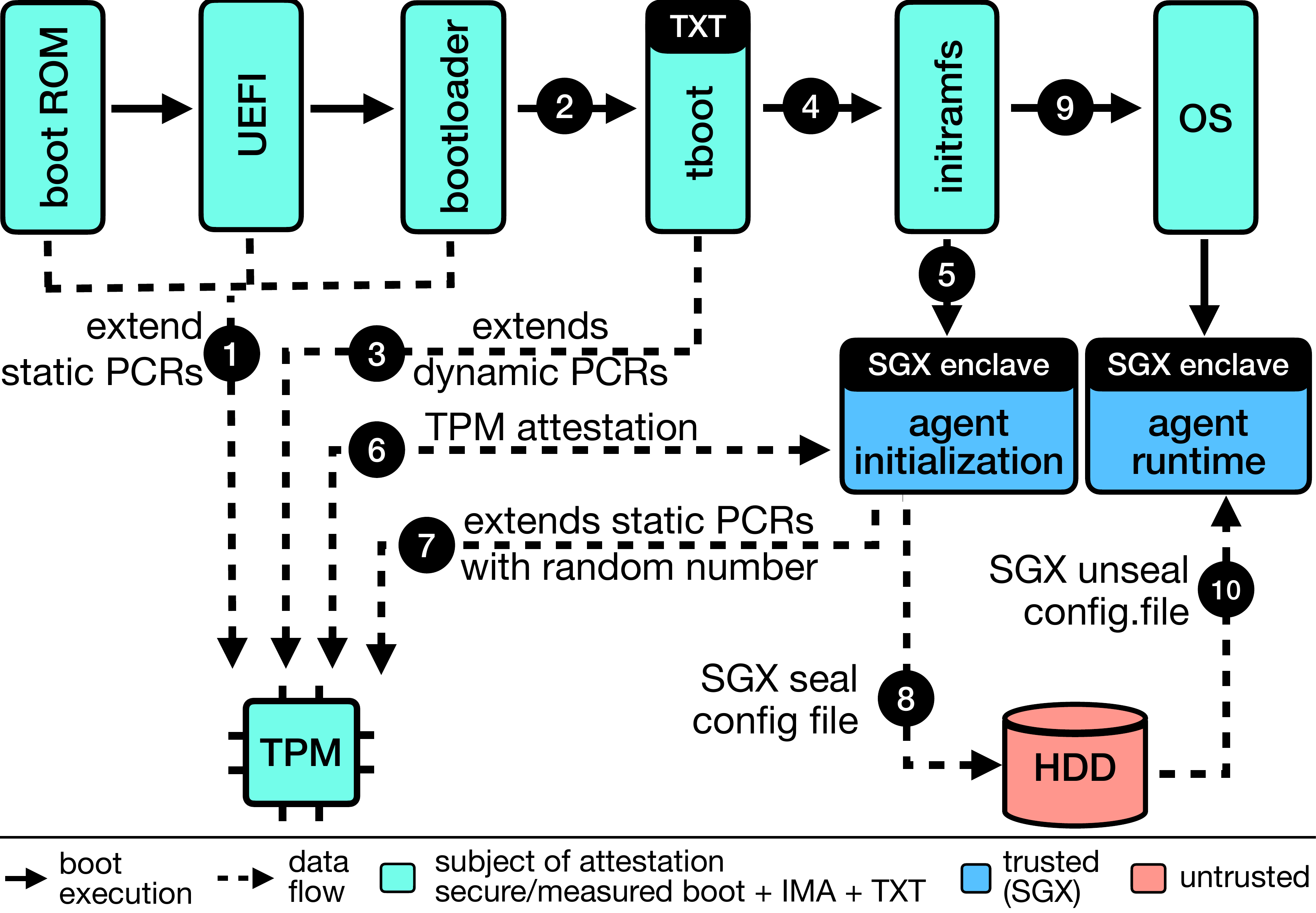}
    \caption{
        The platform boot process. To make the \cuckooattack detectable, the agent executes twice.
        First, in \sysinit, the agent executes in the measured environment where it shares a secret with the TPM. 
        Second, in \sysruntime, the agent establishes trust with the local TPM or detects the \cuckooattack.
    }
\label{fig:TBEBootstrap}
\end{figure}

\autoref{fig:TBEBootstrap} illustrates the bootstrap of a computer where the agent collects information required to detect the \cuckooattack.
Consecutive \gls{uefi} components execute in the chain of trust; their integrity measurements are extended in static \glspl{pcr} (\raisebox{-1pt}{\ding{202}}). 
\gls{uefi} loads the bootloader, which starts the tboot (\raisebox{-1pt}{\ding{203}}). 
The tboot leverages \gls{txt} \cite{intel_txt_whitepaper,intel_txt_spec}--which implements DRTM on Intel CPUs--to establish a trusted environment. 
The tboot measures the integrity of the Linux kernel and initramfs, extends these measurements to dynamic \glspl{pcr} (\raisebox{-1pt}{\ding{204}}), and executes them (\raisebox{-1pt}{\ding{205}}).

The initramfs has two essential properties; its integrity is reflected in dynamic \glspl{pcr}, and failures during initramfs execution prevent machine booting. 
We rely on these properties to verify that the agent completed its execution. 
We refer to the agent execution inside initramfs as \emph{\sysinit} (\raisebox{-1pt}{\ding{206}}).

During the \sysinit, the agent requests the TPM to create a new \gls{aik}, return the TPM's \gls{ek} certificate, and return the \tpmquote certifying \glspl{pcr} (\raisebox{-1pt}{\ding{207}}). 
The agent performs the \texttt{activation of credential} procedure (\cite{arthur_practical_2015} p. 109-111) to verify that the \gls{aik} was created by the TPM, which possesses the private key associated with the \gls{ek} certificate. 
The agent then \emph{obfuscates} static \glspl{pcr} by extending them with a random number generated inside the \gls{sgx} enclave (\raisebox{-1pt}{\ding{208}}). 
To ensure that the obfuscation succeeded and the boot process to continue, the agent reads \glspl{pcr} again and compares them to the expected pre-computed hashes. 
After all, the AIK, the EK certificate, the TPM clock (includes computer reboot counter), and PCRs (original and obfuscated) are persisted in the file system in the \gls{sgx} sealed \emph{configuration file} (\raisebox{-1pt}{\ding{209}}).
The initramfs handles control to the OS (\raisebox{-1pt}{\ding{210}}), after the \sysinit finishes.
The OS executes the agent together with startup services. 
We refer to the agent execution after the OS executes as \emph{\sysruntime}.

\subsection{Establish Trust}
\label{impl:establish_trust}
During the \sysruntime, the agent verifies that there was no \cuckooattack during \sysinit and \sysruntime by ensuring that the following conditions are fulfilled:

\textbf{Condition 1}: the agent is able to unseal the configuration file (\raisebox{-1pt}{\ding{211}}).
Relying on the properties of the SGX unseal, we conclude that the configuration file was created by the agent enclave running the same binary, and both enclaves were executed on the same \gls{sgx} processor.

\textbf{Condition 2}: a successful match between dynamic \glspl{pcr} read from the \gls{tpm} and the golden dynamic \glspl{pcr}.
It proves that during \sysinit, the agent enclave was executed in the trusted environment (Linux kernel, initramfs, and correct TPM driver), and it successfully obfuscated the TPM.

\textbf{Condition 3}: a successful match of static \glspl{pcr} read from the \gls{tpm} with obfuscated static \glspl{pcr} read from the configuration file.
It proves that the configuration file contains the information gathered earlier from the same \gls{tpm}.

\textbf{Condition 4}: a successful match of the reboots counter stored in the configuration with the reboots counter value read from the fresh quote proves that the computer did not reboot since the \sysinit.

Finally, considering conditions 1, 2, 3, 4, and what they indicate once fulfilled, we conclude that the \tpmquote was issued by the TPM that collected software measurements during the computer bootstrap. 
\S\ref{sec:evaluation:formal_analysis} formally proves this claim.

\subsection{Cache Updates}
\label{impl:cache_updates}
To decrease the policy verification latency, the agent starts a separate thread reading the computer state to validate it against future policy verification requests.
The agent recurrently retrieves the \tpmquote and verifies that the quote certifies PCRs values read during the \sysinit, and it repeatedly reads new events from the \imalog. 

Hashes of all events are stored in the enclave's memory, together with the number of bytes read ($\mathcal{B}$), and the last value of IMA PCR ($\mathcal{D}$). 
To read new events, the agent first retrieves the \tpmquote and opens the \imalog file skipping $\mathcal{B}$ bytes.
It then reads a new event from the file and recalculates the integrity hash by extending $\mathcal{D}$ with the event's hash. 
This process is repeated for each new event and finishes when the integrity hash is equal to the hash of the IMA PCR retrieved from the \tpmquote. 
If the agent reaches the end of the \imalog and the integrity hash does not match the hash in the IMA PCR, it detects the tampering of the \imalog and the OS is considered compromised.

\subsection{Policy Verification}
\label{impl:policy_verification}
The agent exposes the policy verification functionality via a TLS-protected \gls{rest} \gls{api} endpoint to simplify the communication interface between verifiers and agents.
It is enough for verifiers to check the agent's identity by verifying its X.509 certificate presented during a TLS-handshake. 
Currently, TLS credentials are delivered to the agent via a \gls{kms}~\cite{palaemon_2020} but the verifier can also rely on the SGX remote attestation~\cite{johnson2016intel} to ensure the agent's identity and integrity.
As future work, the agent will create a self-signed certificate via sgx-ra-tls \cite{intel_sgxra_whitepaper}, thus excluding the \gls{kms} from the trusted computing base.

The agent stores a once deployed policy in the in-memory key-value map under a randomly generated key \emph{policy\_id} to permit tenants to verify the same policy again. 
The agent can be queried with the policy\_id to verify that the OS integrity has not changed since the last verification.
An adversary cannot change once deployed policy because \gls{sgx} protects the agent's memory from tampering, \ie, SGX guarantees integrity, confidentiality, and freshness of data. 

\section{Security Risk Assessment}
\label{eval:risk}
\sys combines different security techniques to build a framework providing technical assurance that applications execute inside \gls{tee} on the trustworthy \gls{os}. However, each technique operates under a different threat model, and a careful analysis of existing attacks is required to claim security guarantees.
 
\subsection{Preventing Physical and Hardware Attacks}
First of all, the applied techniques usually do not protect against hardware and physical attacks. The TPM is vulnerable to simple hardware attacks on the communication bus with the CPU that allows an adversary to reset the TPM~\cite{kauer_oslo_2007}, reply to arbitrary measurements~\cite{bsparks_security_2007}, including measurements corresponding to the DRTM launch~\cite{winter_hijackers_2013}. Similarly, Intel SGX is vulnerable to clock speed and voltage manipulation~\cite{Murdock2019plundervolt}. Direct memory access attacks~\cite{Thunderclap2019} or cold-boot attacks~\cite{halderman2009lest} can compromise the entire operating system and applications that store data in the main memory in plaintext. To prevent these kinds of attacks, we propose to attest to the physical location of the computer. Regulators require that \glspl{dc} are access controlled and place computers inside security cages~\cite{epa}. We argue that these techniques provide enough security to consider physical and hardware attacks inside the trusted data center negligible.

We use the concept of a trusted beacon to verify that the computer is located in the trusted \gls{dc}. In real-world, the trusted beacon functionality could be provided by a hardware security module~\cite{ibmCryptoCard4769} or a trusted timestamping authority running on a computer with formally proved software~\cite{klein_sel4:_2009, protzenko2020evercrypt}. The only assumption is that trusted beacons must be securely placed inside the \gls{dc} and then be protected from being moved. 

\subsection{Establishing Trust with the Agent.}
To verify that the computer is indeed located in the expected \gls{dc}, we must rely on the agent executing on a potentially untrusted computer exposed to physical and hardware attacks. To authenticate the agent and verify that it executes on a genuine Intel SGX CPU, we leverage Intel SGX remote attestation~\cite{johnson2016intel}. In the past, researchers managed to extract Intel SGX attestation keys~\cite{sgaxe, Foreshadow} that allowed impersonating a genuine SGX CPU. The available mitigations are: i) relying on on-premise data center attestation mechanism~\cite{scarlata2018supporting}, ii) checking for revoked SGX attestation keys, and iii) verifying that the agent runs in the proximity of a trusted device to ensure that it is in the correct data center composed of legitimate SGX machines~\cite{dhar2020proximitee}. In all cases, we must trust the CPU manufacturer, SGX design, cryptographic primitives, and CPU implementation. We consider these assumptions practical because they are common industry practices.

\subsection{Establishing Trust with the TPM.}
\sys relies on TXT, SGX, and TPM to detect the cuckoo attack. Researchers demonstrated that malware placed in the \gls{smm} could survive the TXT late launch~\cite{wojtczuk_attacking_2009}. To mitigate attacks on \gls{smm}, Intel introduced an SMI transfer monitor that constrains the system management interrupt handler mitigating these class of attacks entirely. Other TXT-related and tboot vulnerabilities~\cite{wojtczuk_attacking_2011} were related to memory vulnerabilities in Intel's firmware and tboot implementations. 

Intel SGX is vulnerable to microarchitectural and side-channel attacks that violate SGX confidentiality guarantees~\cite{Foreshadow}. Intel constantly patches the vulnerabilities with microcode updates or hardware changes. Nonetheless, we do consider these attacks as a real threat because of their severity and the multitude of variants that appear.

These attacks do not impact \sys guarantees because they only affect SGX confidentiality and not integrity. The only security-sensitive data that might be used to compromise \sys is the secret shared between the agent and the TPM. However, the secret lives only during the \sysinit, where the presence of malware is detected. In more detail, an adversary can extract the secret shared between the agent and the TPM during the \sysinit to mount the cuckoo attack by sharing the secret with an arbitrary TPM. We formally proved (\S\ref{sec:evaluation:formal_analysis}) that the \sys protocol is immune to these kinds of attacks because the agent detects that the secret was leaked once it executes in \sysruntime. The agent detects that malware was present during the \sysinit because both initramfs and kernel are measured by DRTM, and their measurements are securely transferred to the agent in \sysruntime via SGX sealing. An adversary cannot tamper with the sealed data because only the same enclave running on the same CPU can seal and unseal the data. Thus, the presence of malware and secret leakage are revealed. 

\subsection{Establishing Trust with the OS.}
Because the agent can read the load time integrity of the kernel stored inside the dynamic PCR in the TPM, it can ensure that the computer executes a kernel that was intended to load because even if an adversary boots a malicious kernel, she cannot tamper with PCRs that reflect the malicious kernel load. 

An adversary who gains access to the computer by stealing credentials using social engineering or exploiting a misconfiguration cannot run arbitrary software because she does not have the signing key to issue a certificate required by the integrity-enforcement mechanisms (IMA) to authorize the file. 

However, an adversary might exploit memory vulnerabilities in the existing code, such as Linux kernel or software executing on the system remotely~\cite{carvalho2014heartbleed}. This is feasible because most system software is implemented in unsafe memory languages. We assume that the operating system owner relies on an additional security mechanism enumerated in \S\ref{sec:threat_model} to enforce the runtime process integrity. Typically, the system owner also minimizes the \gls{tcb} by authorizing only crucial software to run on a computer. He does it by digitally signing only trusted software and relying on the IMA-appraisal to enforce it during the OS runtime. 

An adversary who gains access to the computer can restart it and disable the security mechanisms or boot the computer into an untrusted state. In \S\ref{eval:performance}, we estimate the vulnerability window size in which the monitoring controller detects the computer integrity violation. 

Another attack vectors are network side-channel attacks, such as NetCAT~\cite{kurth_netcat_2020}, and rowhammer attacks over the network~\cite{tatar2018throwhammer}. In these attacks, an adversary does not have to run malware on the computer but instead sends malicious network packages that modern network cards place directly in the main memory. We assign a low risk to these classes of attacks because i) they are hard to perform in noisy production environment, ii) they are detectable by network traffic monitoring tools and firewalls because they generate high network activity, iii) mitigation techniques exist and can be applied independently~\cite{kurth_netcat_2020, tatar2018throwhammer}.

\section{Evaluation}
\label{sec:evaluation}

We evaluate \sys in four-folds.
In \S\ref{eval:realworld}, we demonstrate \sys protecting a real-world application from the eHealth domain.
Then, in \S\ref{eval:performance} and \S\ref{eval:performance_macro}, we evaluate \sys' security and performance, respectively. 
Finally, in \S\ref{sec:evaluation:formal_analysis} we present the formal verification of the \cuckooattack detection protocol.

\textbf{Testbed.} Experiments execute on a rack-based cluster of three \machinetype servers connected via a 10\,Gb Ethernet.
Each server is equipped with an \machinecpu, 64\,GiB of RAM, \tpmchipversion, running  \ubuntuver with kernel \kernelver. 
The CPUs are on the microcode patch level (\microcodever). The \gls{epc} is configured to reserve 128\,MiB of RAM. 
During all experiments, the agent, the monitoring controller, and the trusted beacon run on different machines.

\newcommand{\YES}{{\ding{51}}}
\newcommand{\NO}{{\ding{55}}}

\begin{table}[b!]
\caption{
    The execution time of the eHealth application.
    Mean values calculated from 30 independent application executions. 
    The standard deviation in all variants was 1\,sec. 
}
\center
\begin{tabular}{lccc}
 & native & SCONE & \sys \\ \hline
Execution time & 41\,sec   & 52\,sec   & 53\,sec   \\
Security level & & & \\
- tolerate rogue operator & \NO & \YES & \YES \\
- tolerate untrusted OS & \NO & \YES & \YES \\
- side-channel attacks & \NO & \NO & \YES \\
- data processed in \\correct geolocation & \NO & \NO & \YES \\
\hline
\end{tabular}
\label{tab:ehealth}
\end{table}

\subsection{Protecting a Real-world eHealth Application}
\label{eval:realworld}
We leveraged \sys to protect an eHealth application provided to us by a partner who requires protection of his intellectual property (the application's source code) and the confidentiality of the privacy-sensitive patients' data. This dataset  contains concentrations of $112$ metabolites in cerebrospinal fluid samples from patients with bacterial meningitis, viral meningitis/encephalitis, and non-inflamed controls. 
The application, implemented in Python, uses a \gls{ml} algorithm to understand pathophysiological networks and mechanisms as well as to identify disease-specific pathways that could serve as targets for host-directed treatments to reduce end-organ damage.
We used publicly available SCONE docker images~\cite{docker_curated_images} to run the application inside a container executed inside the SGX enclave.
We configured the OS to use IMA and run the \sys's agent.
On two other machines, we deployed the trusted beacon and the monitoring controller, which was constantly querying the agent to verify the OS integrity.

We measured the execution time of the machine learning algorithm run in three different variants; 
in \emph{native}, the application executes in the untrusted OS; 
in \emph{SCONE}, the application executes in the untrusted OS but inside an SGX enclave provided by SCONE; 
in \emph{\sys}, the application executes inside the SGX enclave on an integrity-enforced OS booted with \sys.
\autoref{tab:ehealth} shows that the machine learning algorithm's execution inside the SGX enclave takes 52\,sec, which was $1.3\times$ longer than the native execution (41\,sec).
\sys further increased the application execution time by 2\%, compared to the SGX enclave execution.
This is an acceptable performance overhead, assuming the higher security guarantees offered by \sys and the compliance with the privacy regulations required by the EU law.

\subsection{Security}
\label{eval:performance}
An adversary cannot violate the computer system's integrity if all integrity enforcement mechanisms are properly configured and enabled (including mechanisms protecting runtime process integrity \S\ref{sec:threat_model}) because the kernel rejects untrusted files from loading to the memory. 
However, an adversary can run arbitrary software if she gets enough privileges to boot the computer with disabled enforcement mechanisms. 
We run a set of micro-benchmarks to estimate the \emph{vulnerability window size} expressed with Equation (1), during which the integrity violation remains undetected.

\begin{equation}
t_{vw} = t_{rq} + 2*(nt_{re} + t_{vp})
\end{equation}
$t_{vw}$ is the vulnerability window size, $t_{rq}$ is the time to read a TPM quote, $n$ is the maximum number of events that can be opened within $t_{rq}$, $t_{re}$ is the time to read a single event from the \imalog, $t_{vp}$ is the time required by the agent to verify the policy and by a verifier to send, receive, and process the verification request.

\begin{table}[b!]
    \caption{
        The latency of reading the TPM quote generated using different signing schemes.
        Mean values calculated from 30 experiment executions. $\sigma$ stands for standard deviation. 
    }
    \center
    \begin{tabular}{lc}
    Signing scheme & TPM quote read latency \\ \hline
    RSA 2048 with SHA-256 & 521\,ms ($\sigma=4$\,ms) \\
    ECDSA P256 with SHA-256 & 155\,ms ($\sigma=2$\,ms) \\
    HMAC with SHA-256 & 107\,ms ($\sigma=3$\,ms) \\
    \hline
    \end{tabular}
    \label{tab:tpm_quote_read}
\end{table}

\myparagraphnotdot{What is the latency of reading a TPM quote?}
Each time the agent reads the \imalog, it reads a fresh TPM quote to verify the \imalog's integrity.
The TPM supports different signing schemes that have a direct impact on the TPM quote read latency.
\autoref{tab:tpm_quote_read} shows that TPM issues a quote using \gls{hmac} in 107\,ms, which is $4.9\times$ faster than when using \gls{rsa} cryptography and $1.4\times$ faster when using \gls{ecdsa}.
Thus, selecting an \gls{hmac} or \gls{ecdsa} allows validating the \imalog's integrity faster than when using \gls{rsa}.
We assume usage of the \gls{ecdsa} when reading a quote, thus $t_{rq}$=155\,ms.

\begin{table}[b!]
    \caption{
        The latency of reading a single event from the \imalog.
        Mean values calculated from 1200 events readings. $\sigma$ stands for standard deviation. 
    }
    \center
    \begin{tabular}{lc}
     & Read latency of a single \imalog entry \\ \hline
    ImaNg event & 34\,$\mu$s ($\sigma=28$\,$\mu$s) \\
    ImaSig event & 58\,$\mu$s ($\sigma=32$\,$\mu$s) \\
    \hline
    \end{tabular}
    \label{tab:ima_entry}
\end{table}

\myparagraphnotdot{What is the latency of reading integrity measurements?}
We measured the latency of reading new measurements from the \imalog to learn how fast the agent can detect the integrity violation.
During the first read of the \imalog, the agent reads all measurements collected by IMA during the OS boot, which is typically the biggest chunk of the IMA log that has to be read by the agent at once. The bootstrap of Ubuntu Linux produces approximately 1800 measurements. 
The agent needs 130\,ms to read all events from the \imalog, recalculate the \imalog integrity hash, and compare the hash to the \imapcr. 

After the initial \imalog read, the agent reads only the new IMA measurements since the last \imalog read.
The time needed to read the integrity measurements depends on the number of new events measured and added to the \imalog. 
\autoref{tab:ima_entry} shows that the agent requires 34\,$\mu$s and 58\,$\mu$s to retrieve a single ImaNg and ImaSig event, respectively.
The ImaNg, a default IMA event format providing the file's integrity hash.
The ImaSig event entry extends the ImaNg format by also including the file's signature.
So, the maximum event read time $t_{re}$=58\,$\mu$s.

\myparagraphnotdot{How much time does it take to detect the integrity violation?}
\label{eval:vulnerability_window}
The vulnerability window for the attack consists of the time the agent takes to read a fresh quote, retrieve new events from the \imalog, and process the policy verification request.
We assume that when the agent reads a quote ($t_{rq}$), an adversary can cause IMA to open no more than $n$=3875\,files (according to our measures, opening a file takes at least 40\,$\mu$s). The agent would require about $n*t_{re}$=225\,ms to read events, and about $t_{vp}$=100\,ms to verify them against the policy, see \S\ref{eval:performance_macro}.
Therefore, using Equation (1), we estimate that the policy verification protocol has a vulnerability window of approximately $t_{vw}$=805\,ms.

\subsection{Performance}
\label{eval:performance_macro}

\begin{figure}[t!]
    \centering
    \includegraphics[width=0.5\textwidth]{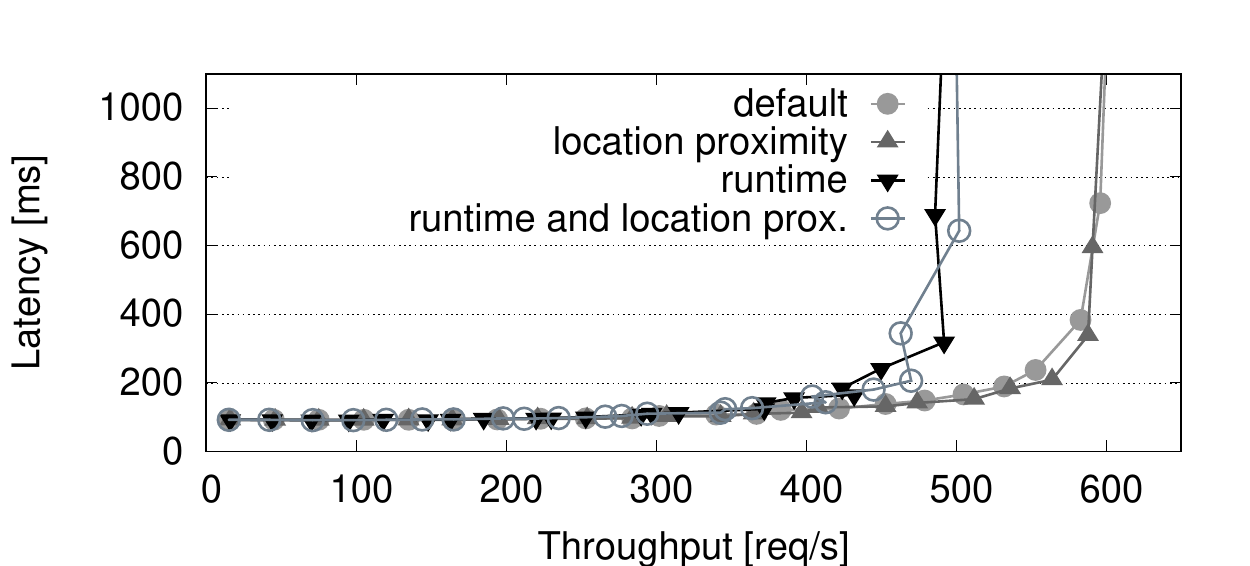}
    \caption{Policy verification throughput. \emph{Default} policy checks secure boot and trusted boot. 
    \emph{Location proximity} checks geolocation. \emph{Runtime} verifies IMA measurements.
    }
    \label{fig:eval:policy_throughput}
\end{figure}

\myparagraphnotdot{How scalable is \sys? Can it efficiently verify policies on behalf of multiple verifiers?}
In our design, the agent is the security-critical component that performs local integrity attestation on behalf of high-assurance security systems, centralized monitoring services, and security officers.  To verify the agent's ability to verify security policies, we measured the policy verification throughput -- the time in which the agent responds to the verifier's request verifying OS integrity. Our experiments compare four variants of the policy content:
i) \emph{default}, the policy contains only the definition of static and dynamic PCRs;
ii) \emph{location proximity}, the default policy content with additional constraints about proximity to trusted beacon;
iii) \emph{runtime}, the default policy content with a whitelist of trusted software;
iv) \emph{runtime and location proximity}, the combination of the runtime and location proximity policies.
\autoref{fig:eval:policy_throughput} shows that the agent achieves the maximum throughput of 623\,req/sec when verifying a default policy. A similar throughput is achieved for the policy with the location proximity extension. The throughput decreases to 521\,req/sec when the agent verifies a security policy containing IMA measurements because of the overhead caused by reading new IMA measurements. 
An optimal latency of 100\,ms is achieved for all policy variants when the throughput $<$ 250\,req/sec.

\begin{table}[b!]
    \caption{
    	The mean remote attestation latency comparison between different integrity monitoring frameworks.
        In all systems, the TPM quote was signed with RSA signing scheme. \emph{se} stands for standard error.
    }
    \center
    \begin{tabular}{lc}
     & Remote attestation latency \\ \hline
    \sys        &   665\,ms (se=2\,ms)          \\
    Intel CIT   &   2475\,ms (se=5\,ms)     \\
    IBM ACS     &   5677\,ms (se=22\,ms)           \\
    \hline
    \end{tabular}
    \label{tab:tbecitacs}
\end{table}

\myparagraphnotdot{How does \sys performance compare to the existing monitoring frameworks?}
We measured the integrity verification latency of the existing integrity monitoring frameworks to check if the presented framework can be considered practical in terms of performance.
Specifically, we compared \sys with \gls{cit}~\cite{opencit_01_org, intel_secl}, and \gls{acs}~\cite{ibm_tpm_acs}, which is a sample code for a \gls{tcg} attestation application. 
We measured the total time taken to establish a connection with an agent, retrieve a fresh quote, and compare \glspl{pcr} with a whitelist.
In all experiments, the TPM has been previously commissioned. 
\autoref{tab:tbecitacs} shows that \sys with the mean latency of 665\,ms outperforms \gls{cit} by $3.7\times$ and \gls{acs} by $8.5\times$. 
\sys achieves better performance because, during the initialization, it caches \gls{aik}, static \glspl{pcr}, and dynamic \glspl{pcr} that do not change during the entire agent's life cycle. The agent verifies that those values did not change by comparing them to the certified values obtained from the \tpmquote. Furthermore, unlike others, the agent verifies the integrity of the \imalog and \glspl{pcr} by recomputing a hash over cached \glspl{pcr} and \imalog and matching it against the \glspl{pcr} hash in the \tpmquote. It allows the agent to skip the slow process of reading PCRs and, consequently, reduce communication with the TPM to a single recurrent \tpmquote read operation. 

\myparagraphnotdot{How much time does it take to deploy a single security policy?}

\begin{table}[b!]
    \caption{
        The latency of the policy deployment into the agent depending on the content of the security policy.
        Mean values calculated from 600 independent policy deployments. $\sigma$ stands for standard deviation. 
    }
    \center
    \begin{tabular}{lc}
    Security policy content & Deployment latency \\ \hline
    Static and dynamic PCRs & 576\,ms ($\sigma=15$\,ms) \\
    + location proximity & 626\,ms ($\sigma=17$\,ms) \\
    + IMA measurements & 606\,ms ($\sigma=16$\,ms) \\
    + location prox. and IMA measur. & 677\,ms ($\sigma=15$\,ms) \\
    \hline
    \end{tabular}
    \label{tab:policy_deployment}
    \end{table}

\autoref{tab:policy_deployment} shows the latency of the policy deployment protocol using different policy extensions. The latency is measured as the total time between establishing a \gls{tls} connection with \sys, a policy upload, a verification using a fresh quote, and a response retrieval.
The default policy's size, containing the whitelist of 13 \glspl{pcr} and one \gls{tpm} manufacturer's \gls{ca} certificate, is 4.7\,kB. Its deployment takes 576\,ms.
The runtime policy size, containing the whitelist of 1790 files and an IMA signing certificate, is 235\,kB ($50\times$ the default policy). Its deployment lasts 606\,ms, which is only a $1.05\times$ of the default policy deployment latency. The deployment latency of a policy with the location proximity extension depends on the communication latency between \sys and trusted beacons. The deployment of the policy with one trusted beacon located in the same data center takes 626\,ms.

\begin{figure}[t!]
    \centering
    \includegraphics[width=0.5\textwidth]{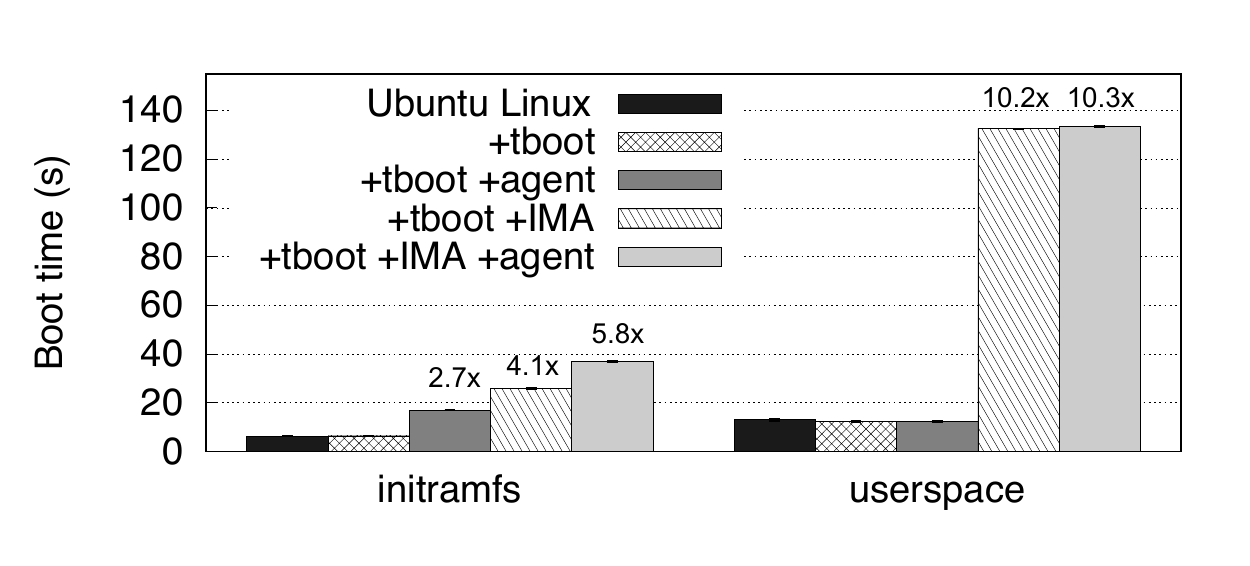}
    \caption{Impact of \sys on boot time.}
    \label{fig:eval:boot_time}
\end{figure}

\myparagraphnotdot{How does \sys impact the boot time of a computer?}
We used the {\em systemd-analyze} tool to measure the load time of \gls{initramfs} and userspace in different configuration variants of Ubuntu. \autoref{fig:eval:boot_time} shows that the native Ubuntu Linux starts in 19\,sec, from which the load of the userspace takes 13\,sec and the kernel with initramfs remaining 6\,sec. tboot executes after the bootloader and before the \gls{initramfs}, thus not influencing the load time of the \gls{os}. The activation of \gls{ima} configured to measure all files defined by the \gls{tcg} group (ima\_tcb boot option), increases the boot time to 158\,sec, $8.3\times$ of the native. A load of userspace takes 84\% of this time, which is caused by the measurement of 1790\,files. The boot time could be decreased by reducing the number of services loaded by the \gls{os}. \sys increases the boot time by 58\% compared to the Ubuntu Linux with tboot and 8\% compared to the Ubuntu Linux with \gls{ima}. The increased boot time is mostly caused by the execution of time-consuming \gls{tpm} operations in \gls{initramfs} performed by \sys and \gls{ima}.

\subsection{Formal Analysis}
\label{sec:evaluation:formal_analysis}

We propose the PCR \emph{obfuscation} as a resilience mechanism against the \cuckooattack.
To prove this claim, we formally verified the protocol's integrity without/with \emph{obfuscation} using the SAPIC tool~\cite{SAPIC}.
SAPIC allows to model security protocols in a variant of applied pi calculus~\cite{Milner1997} that handles parallel processes with a non-monotonic global state needed for a security API such as a TPM.
\textit{The protocol model} describes the actions of agents participating in the protocol, the adversary's specification, and the desired security properties.
The adversary and the protocol interact by sending/receiving messages through the network, which changes the system state and creates traces of state transitions. 
\textit{Security properties} are modeled as trace properties, checked against traces of the transition system, or as an observational equivalence of two transition systems.
While the adversary tries to violate the security properties, she is limited by the constraints of cryptographic primitives.
The SAPIC tool uses constraint solving to perform an exhaustive, symbolic search for executions with satisfying traces.
Since the correctness of security protocols is an undecidable problem, the tool may not terminate on a given verification problem.
If it terminates, it returns either proof that the protocol fulfills the security property or a counterexample representing an attack that violates the stated property.

\myparagraph{Model overview}
\autoref{lst:model-description} shows a high-level overview model of the protocol. It has three processes: \emph{Golden}, \emph{TPM}, and \emph{Machine} (line \ref{line:process}). 
The \emph{TPM} and \emph{Machine} processes execute in parallel without limiting the number of instances.

\lstdefinestyle{interfaces}{
  abovecaptionskip=0mm,
  aboveskip=2mm,
    xleftmargin=0.3cm
}
\begin{lstlisting}[caption= The protocol model without PCR obfuscation,label={lst:model-description},language=sapic,breaklines=true,breakatwhitespace=true,postbreak=\mbox{\textcolor{red}{$\Rdsh$}\space},style=interfaces,escapechar=^, mathescape=true]
/* || : parallel process, ! : replicated process
in(msg), out(msg): send/receive message
pk(priv): public key of key priv
hash(value): one-way hash function
<v1,v2,..>: concatenate values
sign(value, key): signs values with key
verify(value, key): checks the value's signature using key
senc(value, key): symmetric encryption of value using key */ 
Protocol = Golden; (!TPM || !Machine)  ^\label{line:process}^

Golden = ^\label{line:sG}^ // golden hashes are made available to the network 
  new CA_priv;
  out(CA_pub = pk(CA_priv)); //asymm. key pair ^\label{line:cakeys}^ 
  new ^UEFI\_{golden}^; out(sign(^UEFI\_{golden}^, ^CA\_{priv}^)); 
  new ^tboot\_{golden}^; out(sign(^tboot\_{golden}^, ^CA\_{priv}^));
  new ^initramfs\_{golden}^; out(^initramfs\_{golden}^);
  new ^kernel\_{golden}^; out(^kernel\_{golden}^) ^\label{line:eG}^
    
TPM = ^\label{line:sT}^ //creates new TPM, quote can be acquired
  new TPM; new AIK_priv; 
  sPCR_extend(TPM, sPCR); dPCR_extend(TPM, dPCR);
  out(TPM, AIK_pub = pk(AIK_priv)); //public key available
  !create_TPM_quote //replicated

create_TPM_quote =^\label{line:scq}^ 
  in(^nonce^); //prevents replay attacks
  <sPCR,dPCR> = read_local_pcr(TPM);
  quote = sign(<sPCR,dPCR,nonce>,AIK_priv)
  out(quote, AIK_pub)^\label{line:eT}^ // signed quote checked with AIK_pub

Machine = ^\label{line:sM}^ //creates new machine
  new Machine; new seal_key;// machine's CPU seal key
  //attach a specific TPM to this specific machine
  in(TPM_local); // connect to a TPM
  in(UEFI_signed); verify(UEFI_signed, CA_pub); //trusted ^\label{line:ssrtm}^ 
  sPCR_extend(TPM_local, hash(UEFI_signed));// static PCR^\label{line:esrtm}^
  in(tboot_signed);verify(tboot_signed,CA_pub);//trusted ^\label{line:sdrtmS1}^
  in(initramfs);  in(kernel); // might be malicious
  extend_with = hash(<tboot_signed, initramfs, kernel>)
  dPCR_extend(TPM_local, extend_with);// dynamic PCR ^\label{line:edrtmS1}^
  agent_initialization

agent_initialization =  // might be malicious ^\label{line:stbe1}^
  new nonce;  out(nonce); 
  //Golden initramfs contains a driver that
  //guarantees communication with the local TPM  
  if initramfs = initramfs_golden then
        TPM = TPM_local; ^\label{line:stbe1g}^    // trusted, connect to local TPM
  else TPM=in(TPM_remote);//malicious, connect to remote TPM ^\label{line:stbe1m}^^\label{line:etbe1m}^
  <sPCR, dPCR, nonce> = in(quote, AIK_pub(TPM)); // read quote of the connected TPM
  verify(quote, AIK_pub(TPM)); // check the signature 
  seal_value = senc(<sPCR, dPCR, AIK_pub(TPM)>, seal_key); // machine specific CPU seal key ^\label{line:stbe1seal}^
  seal_to_disk(Machine,seal_value);  ^\label{line:etbe1g}^  ^\label{line:etbe1seal}^
  agent_runtime ^\label{line:etbe1}^
  		
agent_runtime = // might be malicious ^\label{line:stbe2}^
  <sPCR1,dPCR1,AIK_pub(TPM_sealed)> = unseal_from_disk(Machine,seal_value); ^\label{line:etbe1guns}^  
  new nonce;  out(nonce);
  <sPCR2, dPCR2, nonce> = in(quote, AIK_pub(TPM_sealed));
  verify(quote,AIK_pub(TPM_sealed)); ^\label{line:etbe1gver}^  
  if equal(sPCR1,sPCR_golden,sPCR2) AND  ^\label{line:sCond}^
     equal (dPCR1,dPCR_golden,dPCR2) ^\label{line:eCond}^
      // trigger event (TPM quote represents local machine state)
  then  event MachineTrusted(TPM_sealed, TPM_local);  ^\label{line:eM}^ ^\label{line:etbe2}^
\end{lstlisting}

The \emph{Golden} process (lines~\ref{line:sG}-\ref{line:eG}) constructs a trusted image that complies with the required policy. It makes the image available for other processes and the adversary via the public network.
A trusted authority keeps a private key (CA\_priv) secure while the public part (CA\_pub) is made available to the public network (line \ref{line:cakeys}).
The handles of the UEFI and the tboot (UEFI\_golden, tboot\_golden) are signed with the private key of the trusted authority to mimic that hardware enforces the boot of the correct software only.
The handles of all boot components are available to the public network.
The \emph{TPM} process (lines \ref{line:sT}-\ref{line:eT}) models a TPM chip with static (sPCR) and dynamic (dPCR) PCRs.
It creates a signed quote using a TPM-specific private key (AIK\_{priv}) repeatedly (lines \ref{line:scq}-\ref{line:eT}).

The Machine process models provisioning of a physical machine (lines \ref{line:sM}-\ref{line:eM}) that has a genuine TPM chip, and a TXT- and SGX-capable CPU.
The boot consists of three steps.

\textit{Step 1}, the \emph{Machine} process gets handles of the (trusted) signed tboot, the unsigned initramfs, and the unsigned kernel. 
It extends dynamic PCRs with measurements of the initramfs and kernel (lines \ref{line:sdrtmS1}-\ref{line:edrtmS1}).
Due to the locality protection, only processes in this step are allowed to extend dynamic PCRs. 
Therefore, PCRs of the attached TPM reflect measurements of the loaded execution environment.
So far, the adversary has no control over the machine (lines \ref{line:sdrtmS1}-\ref{line:edrtmS1}).

\textit{Step 2}, \sysinit executes the already measured and loaded initramfs (lines \ref{line:stbe1}-\ref{line:etbe1}). 
Through the previously measured TPM driver, it requests to contact the TPM.
If the TPM driver is malicious, the adversary might provide an AIK\_pub of the remote TPM (line \ref{line:stbe1m}) instead of the locally attached one (line \ref{line:stbe1g}).
Next, \sysinit reads the quote signed with the TPM private key (AIK\_priv).
The quote is verified using AIK\_pub.
The AIK public key and PCRs are sealed to the disk using the local CPU's SGX sealing key (lines \ref{line:stbe1seal}-\ref{line:etbe1seal}).

\textit{Step 3}, the OS and the TPM driver are untrusted. 
The OS takes over the control, the TPM driver is loaded, and the \sysruntime is executed (lines \ref{line:stbe2}-\ref{line:etbe2}).
After unsealing from the disk (line \ref{line:etbe1guns}), \sysruntime reads the quote through the untrusted TPM driver.
The quote is verified with the unsealed attestation public key (line \ref{line:etbe1gver}).
Finally, \sysruntime reports that the execution environment complies with the policy (line \ref{line:etbe2}) if-and-only-if the unsealed PCRs and the quote PCRs both match the golden PCRs values (lines \ref{line:sCond}-\ref{line:eCond}).

\myparagraph{Security property}
The integrity of the protocol is specified as \textit{"if the TPM quote read by \sysruntime matches the unsealed information, its execution environment MUST correspond to the matched values"}. 
\autoref{lst:model-properties} states this property.

\lstdefinestyle{interfaces}{
  float=t,
  floatplacement=p,
  abovecaptionskip=0mm,
  aboveskip=0mm,
    xleftmargin=0.3cm
}
\begin{lstlisting}[style=interfaces,caption=Security property,label={lst:model-properties},language=sapic,breaklines=true,breakatwhitespace=true,postbreak=\mbox{\textcolor{red}{$\Rdsh$}\space},style=interfaces,escapechar=^, mathescape=true]
// Security property - integrity: 
// the checked state and the local state should match
All x y #i. $\textbf{MachineTrusted(x,y)}$@i $\Rightarrow$  x=y ^\label{line:S}^
\end{lstlisting}
SAPIC tool reported a violation to the given property when using the protocol specified in \autoref{lst:model-description}.
The trace describes that the adversary owns two machines: provisioned 
($\mathcal{M}_{p}$) and oracle ($\mathcal{M}_{o}$), connected to $\mathcal{TPM}_{p}$ and $\mathcal{TPM}_{o}$, respectively.
$\mathcal{M}_{p}$ runs a malicious init\-ramfs and OS (sPCR\_golden, dPCR\_malicious), but uses a genuine hardware (TPM and CPU).
The adversary wants to verify it as a trustworthy machine.
To do so, she forwards the requests to the other machine $\mathcal{M}_{o}$, which runs a trustworthy environment with untampered software and genuine hardware (sPCR\_golden, dPCR\_golden).
Note that forwarding read requests to $\mathcal{M}_{o}$ does not require a change to its environment; however, the adversary cannot extend PCRs of $\mathcal{TPM}_{o}$ attached to $\mathcal{M}_{o}$ without changing initramfs and OS, which consequently would change the corresponding dPCR in $\mathcal{TPM}_{o}$.
During \sysinit, the malicious initramfs in $\mathcal{M}_{p}$ forwards the attestation request to $\mathcal{M}_{o}$, which responds with a signed quote from $\mathcal{TPM}_{o}$ that has the PCRs\_golden and can be verified using AIK\_pub($\mathcal{TPM}_{o}$).
PCR values and AIK\_pub keys are sealed using seal\_key of $\mathcal{M}_{p}$.
During \sysruntime, $\mathcal{M}_{p}$ unseals the values from the disk and contacts $\mathcal{TPM}_{o}$ through the malicious OS to get the quote.
The quote contains PCRs that match both the golden and the sealed PCRs.
So, the event (line \ref{line:eM}) is triggered with unequal TPM\_sealed ($\mathcal{TPM}_{o}$) and TPM\_local ($\mathcal{TPM}_{p}$), which indicates the \cuckooattack.
The vulnerability exists because TPM attestation does not guarantee that the received credentials (AIK\_pub) belong to the attested machine.

\lstdefinestyle{interfaces}{
  float=t,
  floatplacement=t,
  abovecaptionskip=0mm,
  aboveskip=0mm,
    xleftmargin=0.3cm
}
\begin{lstlisting}[firstnumber = 30, caption= The protocol model with PCR obfuscation,label={lst:model-extension},language=sapic,breaklines=true,breakatwhitespace=true,postbreak=\mbox{\textcolor{red}{$\Rdsh$}\space},style=interfaces,escapechar=^, mathescape=true]
Machine =  //creates new machine
  new RND; ... ^\label{line:rnd}^ $\stopnumber$ 
$\startnumber{42}$ 
agent_initialization =
  new nonce1;  out(nonce1);
  if initramfs = initramfs_golden then
        TPM = TPM_local; // connect to local TPM
  else TPM = TPM_remote; // connect to remote TPM 
  <sPCR, dPCR, nonce1> = in(quote, AIK_pub(TPM));
  verify(quote, AIK_pub(TPM)); // check the signature 
  sPCR_extend(TPM, RND)); // share the secret with TPM^\label{line:s_extendspcr}^
  new nonce2;  out(nonce2);//read again to ensure sPCR extended
  <sPCR_obf, dPCR, nonce2> = in(quote, AIK_pub(TPM)); 
  verify(quote, AIK_pub(TPM)); // check the signature  ^\label{line:e_extendspcr}^
  if sPCR_obf = hash(<sPCR, RND>) then ^\label{line:s_sealToDiskRAD}^ 
   seal_value = senc(<sPCR, sPCR_obf, dPCR, AIK_pub(TPM)>, seal_key);
   seal_to_disk(Machine,seal_value);  ^\label{line:e_sealToDiskRAD}^  
   agent_runtime ^\label{line:}^  

agent_runtime = 
  <sPCR1,sPCR1_obf,dPCR1,AIK_pub(TPM_sealed)> = unseal_from_disk(Machine,seal_value);
  new nonce;  out(nonce);
  <sPCR2_obf,dPCR2,nonce> = in(quote,AIK_pub(TPM_sealed));
  verify(quote,AIK_pub(TPM_sealed)); 
  if equal(dPCR1,dPCR_golden,dPCR2) AND ^\label{line:s_trust}^  
     equal(sPCR1_obf,sPCR2_obf) AND
     equal(sPCR1,sPCR_golden)
      // trigger event (TPM quote represents local machine state)
  then  event MachineTrusted(TPM_sealed, TPM_local);  ^\label{line:e_trust}^ 
\end{lstlisting}

\myparagraph{Model extension}
\autoref{lst:model-extension} shows the extended model of the protocol to implement the PCR \emph{obfuscation}. 
It required the following changes:
i) Generation of a random number (RND) (line~\ref{line:rnd});
ii) PCRs obfuscation: the static PCRs are extended with the RND (lines \ref{line:s_extendspcr}-\ref{line:e_extendspcr});
iii) \sysinit seals both the original and obfuscated PCRs (lines \ref{line:s_sealToDiskRAD}-\ref{line:e_sealToDiskRAD});
iv) \sysruntime declares a machine trusted if-and-only-if: 
a) the dynamic PCRs golden, sealed and read from the quote match,
b) the obfuscated static PCRs sealed and read from quote match, and 
c) the original static PCRs golden and sealed match (lines \ref{line:s_trust}-\ref{line:e_trust}).

We checked the model extended with the \emph{obfuscation} (\autoref{lst:model-extension}) against the integrity property in \autoref{lst:model-properties}. 
SAPIC tool terminated and reported that all traces of the protocol preserve the given property.
The modification to the model, where the \sysinit enclave shares a secret with the TPM potentially belonging to the attested machine, overcomes the previously described vulnerability.

\section{Related work}
\label{sec:related_work}

Like the existing monitoring systems~\cite{opencit_01_org, ibm_tpm_acs}, \sys relies on the TPM attestation protocol to verify the computer's integrity. 
Unlike them, \sys is resilient to the cuckoo attack. Existing defenses against this attack have a limited application for high-assurance security systems. Fink et al. proposed a time side-channel approach~\cite{fink_catching_2011} to detect the cuckoo attack. As confirmed by the authors, it is prone to false positives and requires stable measurement conditions, an impractical assumption in real-world scenarios. Flicker~\cite{flicker2008} accesses local TPM from the isolated execution environment established by \gls{drtm}. However, DRTM does not attest to the computer location which makes its attestation untrustworthy due to simple hardware attacks~\cite{winter2013hijacker}. Moreover, DRTM permits executing only a single process on the entire CPU at the same time. This impacts application's throughput because a single context switch to DRTM-established environment takes 10-100s of milliseconds~\cite{mccune_trustvisor:_2010}. \sys instead first verifies that the computer is in the trusted data center (thus, no hardware attacks are possible) and uses DRTM only once when provisioning the TPM. This approach provides better performance as required by modern applications. 

Other solutions for root of trust identification problem require the verifier to solve biometric challenge~\cite{nunes2021root}, observing emmited LED signals~\cite{sun2015trustice}, verifying the device state displayed on the screen~\cite{danisevskis2015graphical, lange2013crossover}, using trusted devices to scan bar codes sealed on the device~\cite{mccune2005seeing}, or pressing a special-purpose button for bootstrapping trust during the computer boot~\cite{parno_bootstrapping}. These approaches have limitations because i) the TPM is a passive device controlled by software which, due to lack of trusted I/O paths to external devices, can redirect, reply, or fool the communication, and ii) they require human interaction and thus do not scale for the \gls{dc}-level. 

Recently, Dhar et al. proposed ProximiTEE~\cite{dhar2020proximitee} to deal with the SGX (\emph{not} TPM) cuckoo attack by attaching a trusted device to the computer and detecting the cuckoo attack during the SGX attestation. This solution can verify that the SGX enclave executes on the computer with the attached trusted device because of the very low communication latency between the enclave and the device. Although, as denoted by Parno~\cite{parno_bootstrapping} this approach cannot be used to detect the TPM cuckoo attack because of the slow speed of the TPM, \sys could use ProximiTEE as a trusted beacon implementation to prove that the computer is located in the expected data center. 

Other work focuses on tolerating malware in the OS while preventing side-channel attacks on TEEs. There are three approaches to mitigate these attacks: i) static vulnerability detection~\cite{guarnieri2020spectector, oleksenko2020specfuzz}, ii) attack prevention~\cite{obfuscuro, DRSGX, ge2019time}, and iii) attack detection~\cite{varys_2018, chen2018racing}. The first one consists of analyzing and modifying source code to detect gadgets~\cite{guarnieri2020spectector, oleksenko2020specfuzz}. However, finding all gadgets is difficult or impossible because the search narrows to gadgets specific to known attacks. The second approach prevents attacks by hiding access patterns using oblivious execution/access pattern obfuscation, resource isolation~\cite{ge2019time}, or hardware changes~\cite{weisse2019nda}. These techniques address only specific attacks~\cite{ge2019time}, require hardware changes~\cite{weisse2019nda}, or incur large performance overhead~\cite{obfuscuro, DRSGX}. The last approach consists of runtime attack detection~\cite{varys_2018, chen2018racing} by isolating and monitoring resources of instrumented programs. But, it targets selected attacks and assumes some amount of statistical misses. \sys aims at preventing such attacks without requiring source code changes or hardware modifications, with low performance overhead but a larger trusted computing base.
\section{Conclusion}
\label{sec:conclusion}

We responded to regulatory demands that require stronger isolation of high-assurance security systems by running them inside trusted execution environments on top of a trustworthy operating system and in the expected geolocation. We demonstrated that the combination of Intel SGX with TPM-based solutions meets such requirements but requires protection against the cuckoo attack. We proposed a novel deterministic defense mechanism against the cuckoo attack and formally proved it. We implemented a framework that monitors and enforces the integrity as well as geolocation of computers running high-assurance security systems and mitigates the cuckoo attack. Our evaluation and security risk assessment show that the \sys is practical.

\bibliographystyle{plain}
\interlinepenalty=10000
\bibliography{paper}

\begin{thebibliography}{10}

\bibitem{obfuscuro}
Adil Ahmad, Byunggill Joe, Yuan Xiao, Yinqian Zhang, Insik Shin, and
  Byoungyoung Lee.
\newblock Obfuscuro: A commodity obfuscation engine on intel sgx.
\newblock In {\em Network and Distributed System Security Symposium}, 2019.

\bibitem{alpine_linux}
{Alpine Linux Development Team}.
\newblock {Alpine Linux - Small. Simple. Secure.}
\newblock \url{https://alpinelinux.org/about/}, accessed on July, 2021.

\bibitem{anati2013innovative}
Ittai Anati, Shay Gueron, Simon Johnson, and Vincent Scarlata.
\newblock Innovative technology for cpu based attestation and sealing.
\newblock In {\em Proceedings of the 2nd international workshop on hardware and
  architectural support for security and privacy}, volume~13, page~7. ACM New
  York, NY, USA, 2013.

\bibitem{arm2009trustzone}
{ARM Limited}.
\newblock Building a secure system using trustzone technology.
\newblock White paper, 2009.

\bibitem{AVISPA}
A.~Armando, D.~Basin, Y.~Boichut, Y.~Chevalier, L.~Compagna, J.~Cuellar,
  P.~Hankes Drielsma, P.~C. He{\'a}m, O.~Kouchnarenko, J.~Mantovani,
  S.~M{\"o}dersheim, D.~von Oheimb, M.~Rusinowitch, J.~Santiago, M.~Turuani,
  L.~Vigan{\`o}, and L.~Vigneron.
\newblock The avispa tool for the automated validation of internet security
  protocols and applications.
\newblock In Kousha Etessami and Sriram~K. Rajamani, editors, {\em Computer
  Aided Verification}, pages 281--285, Berlin, Heidelberg, 2005.

\bibitem{FRWexample1}
Alessandro Armando, Roberto Carbone, Luca Compagna, Jorge Cuellar, and Llanos
  Tobarra.
\newblock {Formal Analysis of SAML 2.0 Web Browser Single Sign-on: Breaking the
  SAML-based Single Sign-on for Google Apps}.
\newblock In {\em {Proceedings of the 6th ACM Workshop on Formal Methods in
  Security Engineering}}, {FMSE '08}, pages {1--10}, {New York, NY, USA},
  {2008}.

\bibitem{arnautov2016scone}
Sergei Arnautov, Bohdan Trach, Franz Gregor, Thomas Knauth, Andre Martin,
  Christian Priebe, Joshua Lind, Divya Muthukumaran, Dan O'Keeffe, Mark
  Stillwell, David Goltzsche, Dave Eyers, R\"udiger Kapitza, Peter Pietzuch,
  and Christof Fetzer.
\newblock {SCONE}: {Secure} linux containers with {Intel} {SGX}.
\newblock In {\em 12th {USENIX} Symposium on Operating Systems Design and
  Implementation ({OSDI} 16)}, pages 689--703, 2016.

\bibitem{arthur_practical_2015}
Will Arthur and David Challener.
\newblock {\em A practical guide to {TPM} 2.0: {Using} the new trusted platform
  module in the new age of security}.
\newblock Springer Nature, 2015.

\bibitem{imasig_updates}
Stefan Berger, Mehmet Kayaalp, Dimitrios Pendarakis, and Mimi Zohar.
\newblock {File Signatures Needed!}
\newblock {\em Linux Plumbers Conference}, 2016.

\bibitem{bhatt2014siem}
Sandeep Bhatt, Pratyusa~K. Manadhata, and Loai Zomlot.
\newblock The operational role of security information and event management
  systems.
\newblock {\em IEEE Security and Privacy (S\&P)}, 12(5):35--41, 2014.

\bibitem{proVerif}
B.~Blanchet.
\newblock An efficient cryptographic protocol verifier based on prolog rules.
\newblock In {\em Proceedings of the 14th IEEE Computer Security Foundations
  Workshop}, pages 82--96, 2001.

\bibitem{FRWexample2}
Matteo Bortolozzo, Matteo Centenaro, Riccardo Focardi, and raham.
\newblock Attacking and fixing pkcs\#11 security tokens.
\newblock In {\em Proceedings of the 17th ACM Conference on Computer and
  Communications Security}, CCS '10, 2010.

\bibitem{DRSGX}
Ferdinand Brasser, Srdjan Capkun, Alexandra Dmitrienko, Tommaso Frassetto, Kari
  Kostiainen, and Ahmad-Reza Sadeghi.
\newblock Dr. sgx: Automated and adjustable side-channel protection for sgx
  using data location randomization.
\newblock In {\em Proceedings of the 35th Annual Computer Security Applications
  Conference}, pages 788--800, 2019.

\bibitem{carvalho2014heartbleed}
Marco Carvalho, Jared DeMott, Richard Ford, and David~A Wheeler.
\newblock Heartbleed 101.
\newblock {\em IEEE security \& privacy}, 12(4):63--67, 2014.

\bibitem{chakrabarti2017intel}
Somnath {Chakrabarti}, Brandon {Baker}, and Mona {Vij}.
\newblock {Intel SGX Enabled Key Manager Service with OpenStack Barbican}.
\newblock {\em arXiv e-prints}, 2017.

\bibitem{simtpm}
Dhiman Chakraborty, Lucjan Hanzlik, and Sven Bugiel.
\newblock simtpm: User-centric {TPM} for mobile devices.
\newblock In {\em 28th {USENIX} Security Symposium ({USENIX} Security 19)},
  pages 533--550. {USENIX} Association, 2019.

\bibitem{chen2018racing}
Guoxing Chen, Wenhao Wang, Tianyu Chen, Sanchuan Chen, Yinqian Zhang, XiaoFeng
  Wang, Ten-Hwang Lai, and Dongdai Lin.
\newblock Racing in hyperspace: Closing hyper-threading side channels on sgx
  with contrived data races.
\newblock In {\em 2018 IEEE Symposium on Security and Privacy (SP)}, pages
  178--194. IEEE, 2018.

\bibitem{costan2016intel}
Victor Costan and Srinivas Devadas.
\newblock Intel sgx explained.
\newblock {\em IACR Cryptol. ePrint Arch.}, 2016(86):1--118, 2016.

\bibitem{danisevskis2015graphical}
Janis Danisevskis, Michael Peter, Jan Nordholz, Matthias Petschick, and Julian
  Vetter.
\newblock Graphical user interface for virtualized mobile handsets.
\newblock {\em IEEE S\&P MoST}, 2015.

\bibitem{nunes2021root}
Ivan De~Oliveira~Nunes, Xuhua Ding, and Gene Tsudik.
\newblock On the root of trust identification problem.
\newblock In {\em Proceedings of the 20th International Conference on
  Information Processing in Sensor Networks (Co-Located with CPS-IoT Week
  2021)}, page 315–327, 2021.

\bibitem{dhar2020proximitee}
Aritra Dhar, Ivan Puddu, Kari Kostiainen, and Srdjan Capkun.
\newblock Proximitee: Hardened sgx attestation by proximity verification.
\newblock In {\em Proceedings of the Tenth ACM Conference on Data and
  Application Security and Privacy}, CODASPY '20, 2020.

\bibitem{DolevYao}
Danny Dolev and Andrew Yao.
\newblock On the security of public key protocols.
\newblock {\em IEEE Transactions on information theory}, 29(2):198--208, 1983.

\bibitem{eperi2021top}
{Eperi}.
\newblock Top tier bank and confidential computing.
\newblock
  \url{https://www.intel.com/content/www/us/en/customer-spotlight/stories/eperi-sgx-customer-story.html},
  accessed on July, 2021.

\bibitem{fink_catching_2011}
Russell~A Fink, Alan~T Sherman, Alexander~O Mitchell, and David~C Challener.
\newblock Catching the cuckoo: {Verifying} tpm proximity using a quote timing
  side-channel.
\newblock In {\em International Conference on Trust and Trustworthy Computing},
  pages 294--301. Springer, 2011.

\bibitem{ge2019time}
Qian Ge, Yuval Yarom, Tom Chothia, and Gernot Heiser.
\newblock Time protection: The missing os abstraction.
\newblock In {\em Proceedings of the Fourteenth EuroSys Conference 2019},
  EuroSys '19, New York, NY, USA, 2019. Association for Computing Machinery.

\bibitem{epa}
{Gematik GmbH}.
\newblock {Systemspezifisches Konzept ePA}.
\newblock
  \url{https://www.vesta-gematik.de/standard/formhandler/324/gemSysL_ePA_V1_3_0.pdf}.

\bibitem{erezept}
{Gematik GmbH}.
\newblock {Systemspezifisches Konzept E-Rezept}.
\newblock
  \url{https://www.vesta-gematik.de/standard/formhandler/324/gemSysL_eRp_V1_0_0_CC6.pdf},
  accessed on July, 2021.

\bibitem{gligor2018establishing}
Virgil Gligor and Maverick Woo.
\newblock Establishing software root of trust unconditionally.
\newblock In {\em Network and Distributed Systems Security (NDSS 2019)}, 2019.

\bibitem{intel_txt_whitepaper}
James Greene.
\newblock Intel trusted execution technology: Hardware-based technology for
  enhancing server platform security.
\newblock {\em Intel Corporation, Copyright}, 2012(8), 2010.

\bibitem{palaemon_2020}
Franz Gregor, Wojciech Ozga, Sebastien Vaucher, Rafael Pires, Do~Le~Quoc,
  Sergei Arnautov, Andre Martin, Valerio Schiavoni, Pascal Felber, and Christof
  Fetzer.
\newblock Trust management as a service: Enabling trusted execution in the face
  of byzantine stakeholders.
\newblock In {\em 2020 50th Annual IEEE/IFIP International Conference on
  Dependable Systems and Networks (DSN)}, pages 502--514. IEEE, 2020.

\bibitem{guarnieri2020spectector}
Marco Guarnieri, Boris K{\"o}pf, Jos{\'e}~F Morales, Jan Reineke, and
  Andr{\'e}s S{\'a}nchez.
\newblock Spectector: principled detection of speculative information flows.
\newblock In {\em 2020 IEEE Symposium on Security and Privacy (SP)}, pages
  1--19. IEEE, 2020.

\bibitem{halderman2009lest}
J~Alex Halderman, Seth~D Schoen, Nadia Heninger, William Clarkson, William
  Paul, Joseph~A Calandrino, Ariel~J Feldman, Jacob Appelbaum, and Edward~W
  Felten.
\newblock Lest we remember: cold-boot attacks on encryption keys.
\newblock {\em Communications of the ACM}, 52(5):91--98, 2009.

\bibitem{ima_appraisal}
Serge Hallyn, Dmitry Kasatkin, David Safford, Reiner Sailer, and M~Zohar.
\newblock {Linux Integrity Measurement Architecture (IMA) - IMA appraisal}.
\newblock \url{https://sourceforge.net/p/linux-ima/wiki/Home/#ima-appraisal},
  accessed on July, 2021.

\bibitem{ibmCryptoCard4769}
{IBM}.
\newblock {IBM} {CEX7S} / 4769 {PCIe} {Cryptographic} {Coprocessor} ({HSM}).
\newblock {\em {IBM} {4769} {Data} {Sheet}}, 2019.

\bibitem{ibm_tpm_acs}
{IBM Corporation}.
\newblock {IBM TPM Attestation Client Server}.
\newblock \url{https://sourceforge.net/projects/ibmtpm20acs/}, accessed on
  July, 2021.

\bibitem{intel_secl}
{Intel}.
\newblock {Intel Security Libraries for Data Center}.
\newblock \url{https://01.org/intel-secl}, accessed on July, 2021.

\bibitem{opencit_01_org}
{Intel and National Security Agency}.
\newblock {Intel Open Cloud Intergrity Technology}.
\newblock \url{https://01.org/opencit}, accessed on July, 2021.

\bibitem{tboot}
{Intel Corporation}.
\newblock {Trusted Boot (tboot)}.
\newblock \url{https://sourceforge.net/projects/tboot/}, accessed on July,
  2021.

\bibitem{intel_txt_spec}
{Intel Corportation}.
\newblock Intel trusted execution techonology--software development guide,
  revision 017.0, 2008.

\bibitem{JayaramMasti_2013}
Ramya Jayaram~Masti, Claudio Marforio, and Srdjan Capkun.
\newblock An architecture for concurrent execution of secure environments in
  clouds.
\newblock In {\em Proceedings of the 2013 ACM workshop on Cloud computing
  security workshop}, pages 11--22, 2013.

\bibitem{johnson2016intel}
Simon Johnson, Vinnie Scarlata, Carlos Rozas, Ernie Brickell, and Frank Mckeen.
\newblock Intel{\textregistered} software guard extensions: Epid provisioning
  and attestation services.
\newblock {\em White Paper}, 1(1-10):119, 2016.

\bibitem{kauer_oslo_2007}
Bernhard Kauer.
\newblock {OSLO}: {Improving} the security of {Trusted} {Computing}.
\newblock {\em USENIX}, 2007.

\bibitem{mustakimur2019origin}
Mustakimur~Rahman Khandaker, Wenqing Liu, Abu Naser, Zhi Wang, and Jie Yang.
\newblock Origin-sensitive control flow integrity.
\newblock In {\em 28th {USENIX} Security Symposium ({USENIX} Security 19)},
  pages 195--211, Santa Clara, CA, August 2019. {USENIX} Association.

\bibitem{klein_sel4:_2009}
Gerwin Klein, Michael Norrish, Thomas Sewell, Harvey Tuch, Simon Winwood, Kevin
  Elphinstone, Gernot Heiser, June Andronick, David Cock, Philip Derrin,
  Dhammika Elkaduwe, Kai Engelhardt, and Rafal Kolanski.
\newblock {seL}4: formal verification of an {OS} kernel.
\newblock In {\em Proceedings of the {ACM} {SIGOPS} 22nd symposium on
  {Operating} systems principles - {SOSP} '09}, Big Sky, Montana, USA, 2009.

\bibitem{intel_sgxra_whitepaper}
Thomas Knauth, Michael Steiner, Somnath Chakrabarti, Li~Lei, Cedric Xing, and
  Mona Vij.
\newblock Integrating remote attestation with transport layer security.
\newblock {\em arXiv preprint arXiv:1801.05863}, 2018.

\bibitem{9095263}
Kari Kostiainen, Aritra Dhar, and Srdjan Capkun.
\newblock Dedicated security chips in the age of secure enclaves.
\newblock {\em IEEE Security and Privacy}, 18(5):38--46, 2020.

\bibitem{SAPIC}
Steve Kremer and Robert Kuennemann.
\newblock Sapic: A stateful applied pi calculus.
\newblock \url{http://sapic.gforge.inria.fr/}, accessed on July, 2021.

\bibitem{fortranix_kms}
Ambuj Kumar, Anand Kashyap, Vinay Phegade, and Jesse Schrater.
\newblock {Self-Defending Key Management Service with Intel SGX}.
\newblock {\em Fortranix Whitepaper}, accessed on July, 2021.

\bibitem{kurth_netcat_2020}
Michael Kurth, Ben Gras, Dennis Andriesse, Cristiano Giuffrida, Herbert Bos,
  and Kaveh Razavi.
\newblock {NetCAT}: {Practical} {Cache} {Attacks} from the {Network}.
\newblock In {\em 2020 IEEE Symposium on Security and Privacy (SP)}, pages
  20--38. IEEE, 2020.

\bibitem{lange2013crossover}
Matthias Lange and Steffen Liebergeld.
\newblock Crossover: secure and usable user interface for mobile devices with
  multiple isolated os personalities.
\newblock In {\em Proceedings of the 29th Annual Computer Security Applications
  Conference}, pages 249--257, 2013.

\bibitem{lee2020keystone}
Dayeol Lee, David Kohlbrenner, Shweta Shinde, Krste Asanovi{\'c}, and Dawn
  Song.
\newblock Keystone: An open framework for architecting trusted execution
  environments.
\newblock In {\em Proceedings of the Fifteenth European Conference on Computer
  Systems ({EuroSys\'20})}, pages 1--16, 2020.

\bibitem{FRWexample3}
Gavin Lowe.
\newblock Breaking and fixing the needham-schroeder public-key protocol using
  fdr.
\newblock In Tiziana Margaria and Bernhard Steffen, editors, {\em Tools and
  Algorithms for the Construction and Analysis of Systems}, 1996.

\bibitem{Thunderclap2019}
A.~Theodore Markettos, Colin Rothwell, Brett~F. Gutstein, Allison Pearce,
  Peter~G. Neumann, Simon~W. Moore, and Robert N.~M. Watson.
\newblock Thunderclap: Exploring vulnerabilities in operating system {IOMMU}
  protection via {DMA} from untrustworthy peripherals.
\newblock In {\em 26th Annual Network and Distributed System Security
  Symposium, {NDSS} 2019, San Diego, California, USA, February 24-27, 2019},
  2019.

\bibitem{matsakis_rust_2014}
Nicholas~D Matsakis and Felix~S Klock.
\newblock The rust language.
\newblock {\em ACM SIGAda Ada Letters}, 34(3):103--104, 2014.

\bibitem{mccune2005seeing}
J.M. McCune, A.~Perrig, and M.K. Reiter.
\newblock Seeing-is-believing: using camera phones for human-verifiable
  authentication.
\newblock In {\em 2005 IEEE Symposium on Security and Privacy (S\&P'05)}, 2005.

\bibitem{mccune_trustvisor:_2010}
Jonathan~M McCune, Yanlin Li, Ning Qu, Zongwei Zhou, Anupam Datta, Virgil
  Gligor, and Adrian Perrig.
\newblock Trustvisor: Efficient tcb reduction and attestation.
\newblock In {\em 2010 IEEE Symposium on Security and Privacy}, pages 143--158.
  IEEE, 2010.

\bibitem{flicker2008}
Jonathan~M McCune, Bryan~J Parno, Adrian Perrig, Michael~K Reiter, and Hiroshi
  Isozaki.
\newblock Flicker: An execution infrastructure for tcb minimization.
\newblock In {\em Proceedings of the 3rd ACM SIGOPS/EuroSys European Conference
  on Computer Systems 2008}, pages 315--328, 2008.

\bibitem{Milner1997}
Robin Milner.
\newblock The pi calculus and its applications.
\newblock In {\em Formal Methods for Open Object-based Distributed Systems},
  pages 3--4. Springer, 1997.

\bibitem{Murdock2019plundervolt}
Kit Murdock, David Oswald, Flavio~D. Garcia, Jo~Van~Bulck, Daniel Gruss, and
  Frank Piessens.
\newblock {Plundervolt}: Software-based fault injection attacks against intel
  sgx.
\newblock In {\em {Proceedings of the 41st IEEE Symposium on Security and
  Privacy (S\&P'20)}}, 2020.

\bibitem{varys_2018}
Oleksii Oleksenko, Bohdan Trach, Robert Krahn, Mark Silberstein, and Christof
  Fetzer.
\newblock Varys: Protecting {SGX} enclaves from practical side-channel attacks.
\newblock In {\em 2018 {Usenix} Annual Technical Conference ({USENIX}{ATC}
  18)}, pages 227--240, 2018.

\bibitem{oleksenko2020specfuzz}
Oleksii Oleksenko, Bohdan Trach, Mark Silberstein, and Christof Fetzer.
\newblock Specfuzz: Bringing spectre-type vulnerabilities to the surface.
\newblock In {\em 29th {USENIX} Security Symposium ({USENIX} Security 20)},
  pages 1481--1498, 2020.

\bibitem{tsr_2020}
Wojciech Ozga, Do~Le Quoc, and Christof Fetzer.
\newblock A practical approach for updating an integrity-enforced operating
  system.
\newblock In {\em Proceedings of the 21st International Middleware Conference},
  pages 311--325, 2020.

\bibitem{parno_bootstrapping}
Bryan Parno.
\newblock Bootstrapping trust in a "trusted" platform.
\newblock In {\em Proceedings of the 3rd Conference on Hot Topics in Security},
  2008.

\bibitem{rootfs_encryption_2005}
Mike Petullo.
\newblock Encrypt your root filesystem.
\newblock {\em Linux Journal}, 2005(129):4, 2005.

\bibitem{protzenko2020evercrypt}
Jonathan Protzenko, Bryan Parno, Aymeric Fromherz, Chris Hawblitzel, Marina
  Polubelova, Karthikeyan Bhargavan, Benjamin Beurdouche, Joonwon Choi, Antoine
  Delignat-Lavaud, C{\'e}dric Fournet, et~al.
\newblock Evercrypt: A fast, verified, cross-platform cryptographic provider.
\newblock In {\em 2020 IEEE Symposium on Security and Privacy (SP)}, pages
  983--1002. IEEE, 2020.

\bibitem{ima_design_2004}
Reiner Sailer, Xiaolan Zhang, Trent Jaeger, and Leendert Van~Doorn.
\newblock Design and implementation of a tcg-based integrity measurement
  architecture.
\newblock In {\em USENIX Security symposium}, volume~13, pages 223--238, 2004.

\bibitem{scarlata2018supporting}
Vinnie Scarlata, Simon Johnson, James Beaney, and Piotr Zmijewski.
\newblock Supporting third party attestation for intel sgx with intel data
  center attestation primitives.
\newblock {\em White paper}, 2018.

\bibitem{docker_curated_images}
{Scontain UG}.
\newblock {SCONE Docker curated images}.
\newblock \url{https://hub.docker.com/u/sconecuratedimages}, accessed on July,
  2021.

\bibitem{seshadri2005pioneer}
Arvind Seshadri, Mark Luk, Elaine Shi, Adrian Perrig, Leendert van Doorn, and
  Pradeep Khosla.
\newblock Pioneer: Verifying code integrity and enforcing untampered code
  execution on legacy systems.
\newblock SOSP '05, 2005.

\bibitem{drtm_tcg}
Jacob Shin, Bill Jacobs, Mark Scott-Nash, Julian Hammersley, Monty Wiseman, Rob
  Spiger, Dick Wilkins, Ralf Findeisen, David Challener, Dalvis Desselle, Steve
  Goodman, Gary Simpson, Kirk Brannock, Amy Nelson, Mark Piwonka, Conan Dailey,
  and Randy Springfield.
\newblock {TCG D-RTM Architecture, Document Version 1.0.0}.
\newblock {\em Trusted Computing Group}, 2013.

\bibitem{bsparks_security_2007}
Evan~R Sparks.
\newblock A {Security} {Assessment} of {Trusted} {Platform} {Modules}.
\newblock {\em Computer Science Technical Report {TR}2007-597}, 2007.

\bibitem{sun2015trustice}
He~Sun, Kun Sun, Yuewu Wang, Jiwu Jing, and Haining Wang.
\newblock Trustice: Hardware-assisted isolated computing environments on mobile
  devices.
\newblock In {\em 2015 45th Annual IEEE/IFIP International Conference on
  Dependable Systems and Networks (DSN '15)}, 2015.

\bibitem{tatar2018throwhammer}
Andrei Tatar, Radhesh~Krishnan Konoth, Elias Athanasopoulos, Cristiano
  Giuffrida, Herbert Bos, and Kaveh Razavi.
\newblock Throwhammer: Rowhammer attacks over the network and defenses.
\newblock In {\em 2018 $\{$USENIX$\}$ Annual Technical Conference
  ($\{$USENIX$\}$$\{$ATC$\}$ 18)}, pages 213--226, 2018.

\bibitem{tcg_ima_spec}
{Trusted Computing Group}.
\newblock {TCG Infrastructure Working Group Architecture Part II - Integrity
  Management, Specification Version 1.0, Revision 1.0}, 2006.

\bibitem{tcg_srtm_bios_spec}
{Trusted Computing Group}.
\newblock {TCG PC Client Specific Implementation Specification for Conventional
  BIOS, Specification Version 1.21, Revision 1.00}, 2012.

\bibitem{tpm_2_0_spec}
{Trusted Computing Group}.
\newblock {TPM Library Specification, Family "2.0", Revision 01.38}, 2016.

\bibitem{tcg_srtm_uefi_spec}
{Trusted Computing Group}.
\newblock {TCG PC Client Platform Firmware Profile Specification, Family 2.0,
  Level 00, Revision 1.04}, 2019.

\bibitem{tcg_tpm_attestation}
{Trusted Computing Group}.
\newblock {TCG Trusted Attestation Protocol (TAP) Information Model for TPM
  Families 1.2 and 2.0 and DICE Family 1.0. Version 1.0, Revision 0.36}, 2019.

\bibitem{Foreshadow}
Jo~Van~Bulck, Marina Minkin, Ofir Weisse, Daniel Genkin, Baris Kasikci, Frank
  Piessens, Mark Silberstein, Thomas~F Wenisch, Yuval Yarom, and Raoul Strackx.
\newblock Foreshadow: Extracting the keys to the intel {SGX} kingdom with
  transient out-of-order execution.
\newblock In {\em 27th {USENIX} Security Symposium ({USENIX} Security 18)},
  pages 991--1008, 2018.

\bibitem{sgaxe}
Stephan van Schaik, Andrew Kwong, Daniel Genkin, and Yuval Yarom.
\newblock {SGAxe}: How {SGX} fails in practice.
\newblock \url{https://sgaxeattack.com/}, 2020.

\bibitem{weisse2019nda}
Ofir Weisse, Ian Neal, Kevin Loughlin, Thomas~F. Wenisch, and Baris Kasikci.
\newblock Nda: Preventing speculative execution attacks at their source.
\newblock In {\em Proceedings of the 52nd Annual IEEE/ACM International
  Symposium on Microarchitecture}, MICRO '52, page 572–586, New York, NY,
  USA, 2019. Association for Computing Machinery.

\bibitem{wilkins_secure_boot_2013}
Richard Wilkins and Brian Richardson.
\newblock {UEFI} secure boot in modern computer security solutions.
\newblock In {\em {UEFI Forum}}, 2013.

\bibitem{winter2013hijacker}
Johannes Winter and Kurt Dietrich.
\newblock A hijacker's guide to communication interfaces of the trusted
  platform module.
\newblock {\em Computers \& Mathematics with Applications}, 2013.

\bibitem{winter_hijackers_2013}
Johannes Winter and Kurt Dietrich.
\newblock A hijacker's guide to communication interfaces of the trusted
  platform module.
\newblock {\em Computers \& Mathematics with Applications}, 2013.

\bibitem{wojtczuk_attacking_2009}
Rafal Wojtczuk and Joanna Rutkowska.
\newblock Attacking {Intel} {Trusted} {Execution} {Technology}.
\newblock In {\em {Black Hat DC}}, {2009}.

\bibitem{wojtczuk_attacking_2011}
Rafal Wojtczuk and Joanna Rutkowska.
\newblock Attacking {Intel} {TXT} via {SINIT} code execution hijacking.
\newblock
  \url{https://invisiblethingslab.com/resources/2011/Attacking_Intel_TXT_via_SINIT_hijacking.pdf},
  accessed on July, 2021.

\bibitem{xu2015controlled}
Yuanzhong Xu, Weidong Cui, and Marcus Peinado.
\newblock Controlled-channel attacks: Deterministic side channels for untrusted
  operating systems.
\newblock In {\em Proceedings of the 2015 IEEE Symposium on Security and
  Privacy}, SP '15, page 640–656, USA, 2015. IEEE Computer Society.

\bibitem{fuzzing}
Andreas Zeller, Rahul Gopinath, Marcel B{\"o}hme, Gordon Fraser, and Christian
  Holler.
\newblock The fuzzing book, 2019.

\bibitem{hacl}
Jean-Karim Zinzindohou{\'e}, Karthikeyan Bhargavan, Jonathan Protzenko, and
  Benjamin Beurdouche.
\newblock Hacl*: A verified modern cryptographic library.
\newblock In {\em Proceedings of the 2017 ACM SIGSAC Conference on Computer and
  Communications Security}, pages 1789--1806, 2017.

\end{thebibliography}

\end{document}